\documentclass[lettersize,journal]{IEEEtran}
\usepackage{amsmath,amsfonts}
\usepackage{algorithmic}
\usepackage{algorithm}
\usepackage{array}
\usepackage[caption=false,font=normalsize,labelfont=sf,textfont=sf]{subfig}
\usepackage{textcomp}
\usepackage{stfloats}
\usepackage{url}
\usepackage{verbatim}
\usepackage{graphicx}
\usepackage{cite}

\usepackage{soul}
\usepackage{tabularx}
\usepackage{makecell}  
\usepackage{booktabs}  
\usepackage{array}     
\usepackage{pifont}
\usepackage{graphicx}    
\usepackage{subfig}  
\usepackage{threeparttable} 
\usepackage{multirow}
\begin{document}

\title{Language of Network: A Generative Pre-trained Model for Encrypted Traffic Comprehension}

\author{Di Zhao, Bo Jiang, Song Liu, Susu Cui, Meng Shen,~\IEEEmembership{Member,~IEEE,} Dongqi Han, Xingmao Guan and Zhigang Lu
\thanks{This research is supported by National Key Research and Development Program of China (No.2023YFC2206402), and the Strategic Priority Research Program of the Chinese Academy of Sciences (No.XDA0460100). This work is also supported by the Program of Key Laboratory of Network Assessment Technology, the Chinese Academy of Sciences, Program of Beijing Key Laboratory of Network Security and Protection Technology.\textit{(Corresponding author: Susu Cui.)}}
\thanks{Di Zhao, Bo Jiang, Song Liu, Susu Cui, Xingmao Guan and Zhigang Lu are with the Institute of Information Engineering, Chinese Academy of Sciences, Beijing, China, and also with the School of Cyber Security, University of Chinese Academy of Sciences, Beijing, China (e-mail: zhaodi@iie.ac.cn, jiangbo@iie.ac.cn, liusong1106@iie.ac.cn, cuisusu@iie.ac.cn, guanxingmao@iie.ac.cn, luzhigang@iie.ac.cn). }
\thanks{Meng Shen is with the School of Cyberspace Security, Beijing Institute of Technology, Beijing, China (e-mail: shenmeng@bit.edu.cn).}
\thanks{Dongqi Han is with the Beijing University of Posts and Telecommunications, Beijing, China (e-mail: handongqi@bupt.edu.cn).}}

\markboth{Journal of \LaTeX\ Class Files,~Vol.~14, No.~8, August~2021}%
{Shell \MakeLowercase{\textit{et al.}}: A Sample Article Using IEEEtran.cls for IEEE Journals}

\IEEEpubid{0000--0000/00\$00.00~\copyright~2021 IEEE}

\maketitle

\begin{abstract}
The increasing demand for privacy protection and security considerations leads to a significant rise in the proportion of encrypted network traffic. Since traffic content becomes unrecognizable after encryption, accurate analysis is challenging, making it difficult to classify applications and detect attacks. Deep learning is currently the predominant approach for encrypted traffic classification through feature analysis. However, these methods face limitations due to their high dependence on labeled data and difficulties in detecting attack variants. First, their performance is highly sensitive to data quality, where the high-cost manual labeling process and dataset imbalance significantly degrade results. Second, the rapid evolution of attack patterns makes it challenging for models to identify new types of attacks. To tackle these challenges, we present GBC, a generative model based on pre-training for encrypted traffic comprehension. Since traditional tokenization methods are primarily designed for natural language, we propose a protocol-aware tokenization approach for encrypted traffic that improves model comprehension of fields specific to network traffic. In addition, GBC employs pre-training to learn general representations from extensive unlabeled traffic data. Through prompt learning, it effectively adapts to various downstream tasks, enabling both high-quality traffic generation and effective detection. Evaluations across multiple datasets demonstrate that GBC achieves superior results in both traffic classification and generation tasks, resulting in a 5\% improvement in F1 score compared to state-of-the-art methods for classification tasks.
\end{abstract}


\section{Introduction}
With the rapid advancement of Internet technology, the demand for secure communication and privacy protection increases significantly, resulting in a growing proportion of encrypted traffic within networks. However, encryption protocols also become a concerning double-edged sword, enabling attackers to conceal their activities, thereby complicating the task for network administrators to promptly identify malicious traffic. Google Transparency Report\cite{GoogleTransparency} shows that nearly 95\% of web traffic is delivered over HTTPS. According to Zscaler's 2024 report on encrypted attacks\cite{Zscaler2024}, 87.2\% of threats they block are transmitted through encrypted channels. This underscores the critical importance of encrypted network traffic identification.

Since traffic features are obfuscated after encryption, early payload-based identification methods gradually become ineffective \cite{bernailleEarlyRecognitionEncrypted2007}. Meanwhile, signature-based approaches have limited application scenarios due to their high dependence on predefined static rules\cite{guptaCategoricalSurveyStateoftheart2020, chibaNewestCollaborativeHybrid2019, dongMBTreeDetectingEncryption2021}. In response to this challenge, researchers propose methods based on traffic behavioral characteristics and statistical features\cite{andersonIdentifyingEncryptedMalware2016, fengCanStillObserve2022, isingizweAnalyzingLearningbasedEncrypted2021, vuRealTimeEvaluationFramework2022, theofanousFingerprintingShadowsUnmasking2024}. These methods focus primarily on distinctive attributes in the connection process that remain observable despite encryption, such as packet size distribution, inter-arrival times, and flow duration. By integrating these features with machine learning algorithms, researchers successfully develop robust methodologies that effectively classify encrypted network traffic. However, machine learning-based methods still have several limitations. First, traditional machine learning approaches such as support vector machine (SVM) and random forest (RF) heavily rely on manually designed features, requiring specialized expert knowledge for feature engineering, which is time-consuming and highly subjective. Second, these conventional methods struggle to capture the complex non-linear relationships inherent in traffic data. Finally, these predefined features may prove insufficient when faced with rapidly evolving network environments or constantly changing application behaviors, limiting the models' ability to generalize across diverse traffic scenarios.

\IEEEpubidadjcol

Driven by these limitations, researchers turn to deep learning methods. With its capability for automatic feature extraction, deep learning models can learn hierarchical feature representations directly from raw data, thereby reducing the reliance on manual intervention\cite{wangDetectingAndroidMalware2018, fengUnmaskingInternetSurvey2025, zhaoERNNErrorResilientRNN2023, dengNetNoiseNetwork2024}. Through multiple layers of non-linear transformations, these models can better model complex traffic patterns and capture hidden data relationships. Although these models offer advantages, their performance heavily depends on the quantity and quality of labeled training data. As a result, this directly affects their generalization ability. The acquisition of such high-quality labeled data faces challenges. Existing public datasets often suffer from outdated samples and limited categories, failing to capture the evolving traffic patterns. Collecting and professionally labeling new real-world traffic data requires significant time and effort. Moreover, the process is complicated by privacy concerns and technical challenges. Such data-related issues severely influence the classification performance of deep learning models, leading to poor detection on emerging attacks and frequent false alarms when encountering traffic patterns that deviate from the training data distribution.

Recent studies demonstrate the successful application of pre-trained models, originally developed for natural language processing and computer vision, to traffic analysis domains as a partial solution to data-related challenges\cite{devlinBERTPretrainingDeep2019}. These models excel at learning generalized representations from large-scale unlabeled data, thus reducing dependency on labeled data while adapting to various downstream tasks\cite{linBERTContextualizedDatagram2022, quTrafficGPTBreakingToken2024, luEfficientlyAdaptingTraffic2024, yuNovelApproachApplication2024}. However, existing pre-trained models in traffic analysis face challenges in their design. Recent advances in this direction often process network traffic by treating packets as simple hexadecimal strings. This general input representation fails to capture the inherent organization of network protocols and packet formats into individual fields. For instance, the version number in TLS protocol (0301) may be split into separate tokens (xx03 01xx) by such byte-level tokenization, thus breaking its original structure. Beyond this fundamental design issue, these models also struggle with generalization when applied to downstream tasks with highly imbalanced data distributions. In traffic classification scenarios, attack samples often represent only a small portion of the training data. As a result, the learned representations tend to favor the dominant patterns found in normal traffic, which may lead to a bias that overlooks the features of malicious activities. This inherent bias may become more pronounced during fine-tuning on such imbalanced datasets\cite{zhangEnhancedFewShotMalware2024}.

To tackle these challenges, we present GBC, a Generative model Based on pre-training for encrypted traffic Comprehension, which excels at both accurate traffic classification and high-fidelity traffic generation. Through a unified pre-training approach, GBC not only effectively learns robust  feature representations that are critical for identifying malicious traffic, but also leverages its advanced generation capabilities to strengthen traffic classification. Building on a novel protocol-aware tokenization scheme, the model creates syntax-constrained synthetic samples that address data imbalance, which in turn enables more precise and comprehensive traffic analysis through structured representation.

Specifically, to address the limitations of byte-level tokenization in existing pre-trained models, we propose a protocol-aware field-level tokenization method to transform raw traffic into structured text-like sequences. Following protocol specifications, we carefully segment traffic into semantically meaningful tokens based on protocol fields. This approach not only preserves the natural structure of network traffic but also effectively leverages the strengths of pre-trained models in handling structured text.

While our protocol-aware tokenization scheme provides a solid foundation for traffic representation, the challenge of data imbalance in traffic analysis remains to be addressed. To tackle this issue, we introduce a syntax-guided traffic generation mechanism that particularly focuses on augmenting minority samples. By synthesizing protocol-compliant traffic, our method can augment training datasets, allowing models to generalize more effectively to novel encryption techniques and previously unseen attack scenarios. Crucially, unlike existing generation approaches that often produce syntactically invalid outputs or are limited to specific fields (e.g., TCP ports) \cite{kholghPACGPTNovelApproach2023, mengNetGPTGenerativePretrained2023, bikmukhamedovGenerativeTransformerFramework2020}, our model fully captures the underlying traffic structures and generates parsable pcap packets to better assist in traffic analysis.

Our contribution can be summarized as follows:
\begin{itemize}
	\item We develop GBC, an efficient pre-trained traffic comprehension model that achieves precise classification performance while generating high-quality network traffic, offering a framework that enhances both malicious traffic recognition and synthetic data quality for cybersecurity.
	\item We propose a structured tokenization strategy that segments raw traffic into protocol-compliant semantic units, effectively preserving the inherent structure of network traffic and better leveraging the capabilities of pre-trained models.
	\item We implement an effective approach for generating synthetic network traffic samples, enabling data augmentation in imbalanced scenarios while maintaining structural integrity.
	\item We experimentally validate the effectiveness of our model. For classification tasks, the model achieves a 5\% improvement in F1-score over state-of-the-art methods. For generation tasks, our synthetic traffic samples prove to be effective for data augmentation, enhancing malicious traffic detection by 9\% in F1-score.
\end{itemize}

The structure of this paper is organized as follows: Section II discusses related works and their limitations, offering additional insights into the current state of research. Section III presents the design of GBC. Section IV details the experiments and evaluations, providing a comprehensive analysis of the obtained results. Finally, Section V concludes the paper and outlines directions for future research.

\section{RELATED WORK}

This section provides an overview of the existing literature related to our work, focusing on network traffic classification and traffic generation. 

\subsection{Traffic Classification}

With the growth and complexity of network traffic, traffic classification plays a crucial role in areas such as network security, bandwidth management and application optimization. Machine learning-based approaches extract statistical features from network traffic for classification, yet face significant limitations. Their feature engineering heavily depends on domain expertise, while extracted features often demonstrate poor transferability across varied operational contexts\cite{liuDistanceBasedMethodBuilding2019, houHandlingLabeledData2022, fuRealtimeRobustMalicious2021, hollandNewDirectionsAutomated2021}. Deep learning-based approaches leverage deep neural networks to automatically learn feature representations from raw traffic or optimize manually extracted features. However, these models depend on large amounts of high-quality labeled data for training and exhibit vulnerability to adversarial samples, creating potential for evasion by strategically designed malicious traffic\cite{liuFSNetFlowSequence2019, bazuhairDetectingMalignEncrypted2020, lotfollahiDeepPacketNovel2020, linEfficientApproachEncrypted2022, huohEncryptedNetworkTraffic2021}.

In recent years, pre-trained models achieve remarkable results in the fields of natural language processing and computer vision, and thus are gradually introduced into the research of network traffic classification. He et al.\cite{hePERTPayloadEncoding2020a} use dynamic word embedding techniques to extract features from encrypted network traffic and pre-train a Transformer-based classifier. Lin et al. \cite{linBERTContextualizedDatagram2022} further innovate by extracting bursts from traffic and designing two self-supervised pre-training tasks. Wang et al.\cite{wangLensFoundationModel2024} constructs Lens, a foundation model based on the T5 architecture, enhancing pattern recognition through three carefully designed pre-training tasks.

In data representation and feature encoding, Zhao et al.\cite{zhaoAnotherTrafficClassifier2023} formats raw traffic data into two-dimensional matrices to capture multi-level information, combining Masked Autoencoders with Transformer architectures for traffic classification. Ferrag et al.\cite{ferragRevolutionizingCyberThreat2024} introduces a privacy-preserving encoding technique called PPFLE, which extracts statistical features from network traffic. By combining it with a Byte Pair Encoding tokenizer, the method ensures data privacy while maintaining efficient feature representation.

Application-specific optimization research also emerges. Manocchio et al.\cite{manocchioFlowTransformerTransformerFramework2024} evaluate the impact of Transformer architectures on flow-based Network Intrusion Detection Systems and introduce the FlowTransformer framework, enabling flexible component replacement and dataset evaluation. Lin et al.\cite{linNovelMultimodalDeep2023} propose the PEAN framework, which leverages unsupervised pre-training to extract features from byte content and length sequences, achieving high-performance traffic classification despite challenges with coarse-grained traffic data.

While these approaches show promising capabilities in feature extraction and classification accuracy enhancement, some fundamental limitations remain unresolved. Most studies treat network traffic merely as simple feature sequences or raw byte streams, overlooking its inherent structural and contextual characteristics. Furthermore, the aforementioned methods rely heavily on large volumes of high-quality labeled data, creating substantial barriers for real-world deployment. This dependency raises serious concerns about model generalizability in environments with limited labeled data or rapidly evolving traffic patterns. The computational complexity of these models also presents practical implementation challenges for traffic analysis, particularly in resource-constrained network devices. These limitations collectively underscore the urgent need for more efficient and adaptable traffic classification approaches that can maintain high accuracy while reducing reliance on extensive data resources and capturing the intrinsic structure of network protocols.

\subsection{Traffic Generation}

Traffic generation plays a crucial role in network security. For example, researchers generate traffic that simulates real network environments to create workloads for evaluating the effectiveness of security measures. In addition, models for traffic analysis also rely on large-scale, high-quality training data. However, collecting such data is challenging due to privacy concerns and the scarcity of new attack samples. In this context, traffic generation technology offers an innovative solution to data sourcing challenges for model training. This method not only generates synthetic data, enabling models to learn more diverse and complex network behavior patterns, but also effectively mitigates data privacy concerns.

Early traffic generation controlled traffic characteristics through manual configuration and rule settings. Most research \cite{adelekeNetworkTrafficGeneration2022, emmerichMindGapComparison2017, kolahiPerformanceMonitoringVarious2011, patilSurveySyntheticTraffic2016} focuses on evaluating the performance of different generators in producing high-throughput, low-latency traffic. To address the limitations of manual configuration, Sommers and Barford\cite{sommersSelfconfiguringNetworkTraffic2004} propose a tool to automatically extract parameters from standard Netflow logs or packet traces.

With the continuous advancement of technology, machine learning and deep learning methods become mainstream. These models learn from real network traffic, extracting characteristics and generating similar synthetic traffic. A typical example is the Generative Adversarial Network (GAN), which has become a mainstream approach for network traffic generation\cite{nukavarapuMirageNetGANbasedFramework2022, chengPACGANPacketGeneration2019, linIDSGANGenerativeAdversarial2022, wangPacketCGANExploratoryStudy2020}. GANs consist of generator and discriminator components that continuously improve the quality and authenticity of generated traffic through adversarial training.

Beyond GANs, researchers explore alternative approaches. Du et al.\cite{duDBWECorbatBackgroundNetwork2023} extract temporal-spatial features to guide generation, while Jiang et al.\cite{jiangNetDiffusionNetworkData2023} apply diffusion models by converting traffic into images and back.

With the rise of large-scale pre-trained language models, their excellent contextual understanding and generation capabilities prompt researchers to explore the possibility of applying them to network traffic generation. The core idea is to consider network traffic as a special kind of language or sequence, leveraging the sequence modeling capabilities of language models to generate high-quality network traffic data.

Bikmukhamedov and Nadeev\cite{bikmukhamedovMultiClassNetworkTraffic2021} use packet size and inter-arrival time as input features, employing GPT-2 to generate packet sequences matching these feature distributions. Kholgh and Kostakos\cite{kholghPACGPTNovelApproach2023} transform input to flow descriptions and generate corresponding Python code with GPT-3 to produce replayable pcap files. The framework only supports simple packets and struggles with complex protocols like DNS. Meng et al.\cite{mengNetGPTGenerativePretrained2023} convert each byte in a packet into its corresponding hexadecimal number to generate tokens and train a GPT-2 model. By adding header field prompts to the input, the model is guided to generate traffic for specific tasks. Qu et al.\cite{quTrafficGPTBreakingToken2024} enhance packet tokenization and utilize a discriminator model to distinguish between real and generated data.

These approaches exhibit promising potential in generating diverse network traffic. However, early generators are relatively simple, offering limited flexibility and producing traffic that lacks authenticity, making it easy for detection systems to identify anomalies. Advanced methods leveraging deep learning and pre-trained models can capture complex patterns, generating more realistic synthetic traffic. Yet these approaches face a fundamental paradox: they demand extensive high-quality training data, while the generation tasks themselves typically aim to address data scarcity. Most techniques can only generate specific feature sequences, limiting their practical applications. Though some methods can produce complete traffic, they frequently lack robust validation mechanisms and fail to replicate the complex behavioral patterns inherent in real network environments, resulting in suboptimal performance. Additionally, these methods generally lack comprehensive verification frameworks for their generated results.

\section{MODEL DESIGN}

In this section, we introduce the architecture of GBC, a model specifically designed to analyze network traffic packets by utilizing protocol-aware tokenization strategies and syntax-guided generation mechanisms.

\subsection{Overview}

\begin{figure*}[!t]
	\centering{\includegraphics[width=\textwidth]{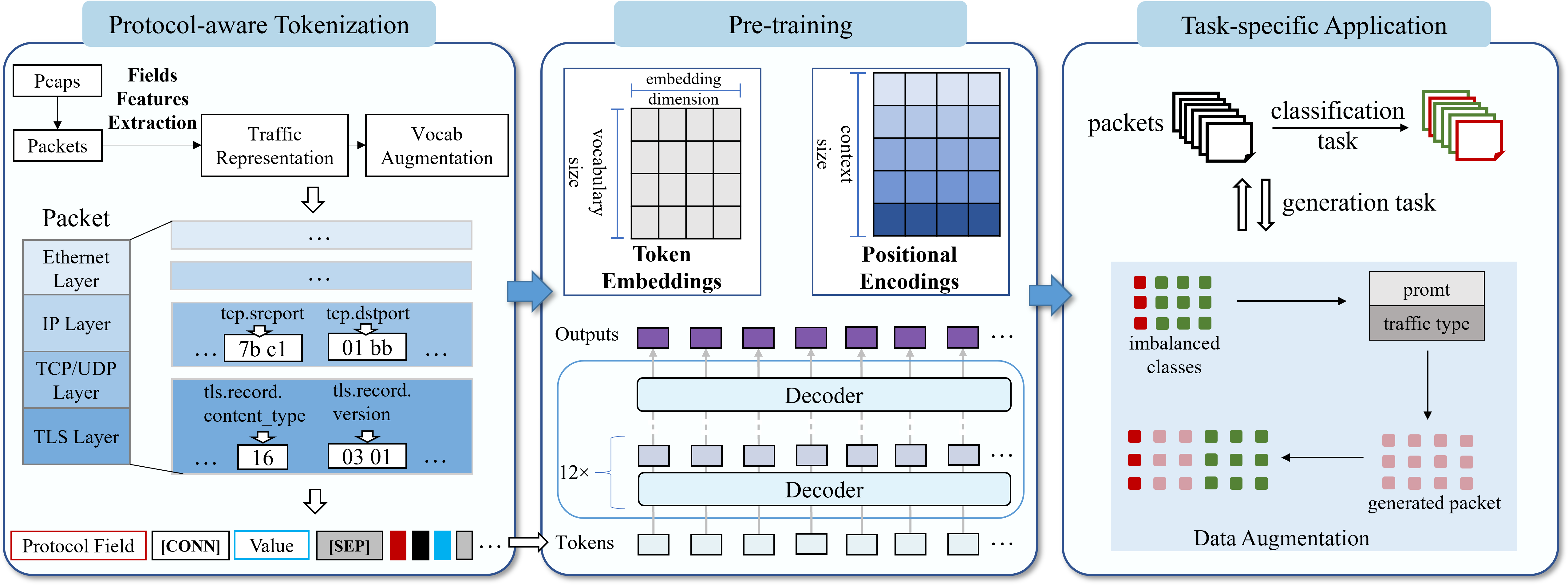}}
	\caption{The framework of GBC. In the preprocessing phase, the model receives network traffic as input, then performs tokenization based on packet structure and network protocol specifications. Following preprocessing, the resulting token sequence serves as input for the next step. The model is pre-trained on large amounts of unlabeled data and fine-tuned for specific tasks using labeled data. When applied to downstream tasks, the model can achieve efficient traffic classification. Additionally, to address issues such as sample imbalance, the model can generate highly realistic traffic for data augmentation, thereby improving performance.}
	\label{fig:all}
\end{figure*}

As shown in Figure \ref{fig:all}, the model is structured to effectively learn general traffic representations, enabling it to perform both encrypted traffic generation and traffic classification. The architecture is divided into three primary components: traffic tokenization, pre-training, and fine-tuning, each of which plays a vital role in enhancing the model's ability to handle complex network traffic analysis tasks. 

During the protocol-aware tokenization phase, network traffic is first segmented into individual packets, with content extracted according to protocol fields. This extraction ensures that the semantic units of the network traffic are preserved, allowing the model to maintain a clear understanding of the underlying structure. The extracted content is then transformed into token representations, which not only preserve the original data but also maintain the hierarchical and structural relationships inherent in network protocols. 

During the pre-training phase, the model is exposed to a vast amount of unlabeled traffic data. Using self-supervised learning, the model learns to capture protocol syntax patterns by predicting parts of the traffic sequence from other observed parts. This unsupervised training on large-scale data allows the model to build a generalized understanding of various network protocols, establishing a robust foundation for both traffic classification and generation tasks. The pre-training phase is crucial for enabling the model to recognize common patterns and features that are not explicitly labeled in real-world traffic data.

Finally, during task-specific application phase, the model is adapted to specific downstream tasks, such as traffic classification or traffic generation. This strategic adaptation phase allows for task-specific optimization while maintaining the foundational knowledge learned during pre-training. This multi-phase approach not only enhances the model’s performance in specialized tasks but also enables it to handle complex network behaviors more effectively.

In the remainder of this section, we will provide a detailed discussion of the design and functionality of each component, illustrating how each stage contributes to the overall performance of the GBC model.

\subsection{Protocol-aware Tokenization}

In this paper, we introduce a protocol-aware tokenization method that explicitly integrates network protocol syntax into the traffic representation. By extracting traffic data and mapping it into a sequence of tokens, this method ensures that the data can be accurately converted back to its original form. It preserves both semantic integrity and protocol structure, which is essential for generating realistic network traffic. Moreover, this approach enables the generation of traffic that can be stored in the standardized pcap format, a widely used format in network traffic analysis, thus facilitating its use in various generation tasks.

Specifically, in order to transform network traffic data into a format that can be processed by the model, it must be tokenized. In the field of natural language processing (NLP), methods such as Byte Pair Encoding (BPE) and WordPiece are commonly used to generate vocabularies. The input text is then segmented and mapped to indices in the vocabulary, resulting in a sequence of tokens that the model can handle. However, network traffic differs significantly from text sequences, thus requiring a more specialized processing method to capture its inherent structure.

For the purpose of uniform processing, current large models for network traffic \cite{linBERTContextualizedDatagram2022, mengNetGPTGenerativePretrained2023, quTrafficGPTBreakingToken2024} always convert each byte of traffic into its corresponding hexadecimal value, treat every two adjacent bytes as a "word", and then use a tokenizer to generate tokens. However, network traffic consists of protocol headers from different layers, exhibiting a clear hierarchical structure. Each field within the protocol headers is specified with corresponding formats, meanings, and value ranges. The aforementioned segmentation strategy may cause structural issues at multiple levels.
\begin{itemize}
	\item Field fragmentation: Fixed-length splitting disregards the natural protocol-defined field boundaries. When a field contains more than two bytes, it is divided into multiple separate tokens, disrupting the inherent protocol structure.
	\item Semantic dissociation: When protocol fields are split into multiple tokens, the semantic meaning of those fields becomes dispersed across different tokens, making it difficult for the model to accurately interpret the original meaning of the data.
	\item Hierarchy loss: The natural layered architecture of network protocols is flattened into a simple sequence of tokens. This loss of hierarchy prevents the model from learning crucial cross-layer interaction patterns and contextual relationships, which are essential for accurate traffic analysis.
\end{itemize}

To overcome these limitations, we propose a network protocol-based traffic representation method that aims to enhance the semantic understanding of network traffic, while fully preserving the original structure of the protocols. This method ensures that the intricate details of network traffic, including its hierarchical structure, are accurately captured and maintained. Furthermore, the specialized terminology and structure inherent in network traffic make the standard GPT-2 vocabulary insufficient for interpreting the detailed nature of traffic data. This limitation necessitates the expansion of the vocabulary to incorporate domain-specific terms and field values.

\begin{figure}[!t]
	\centering{\includegraphics[width=\columnwidth]{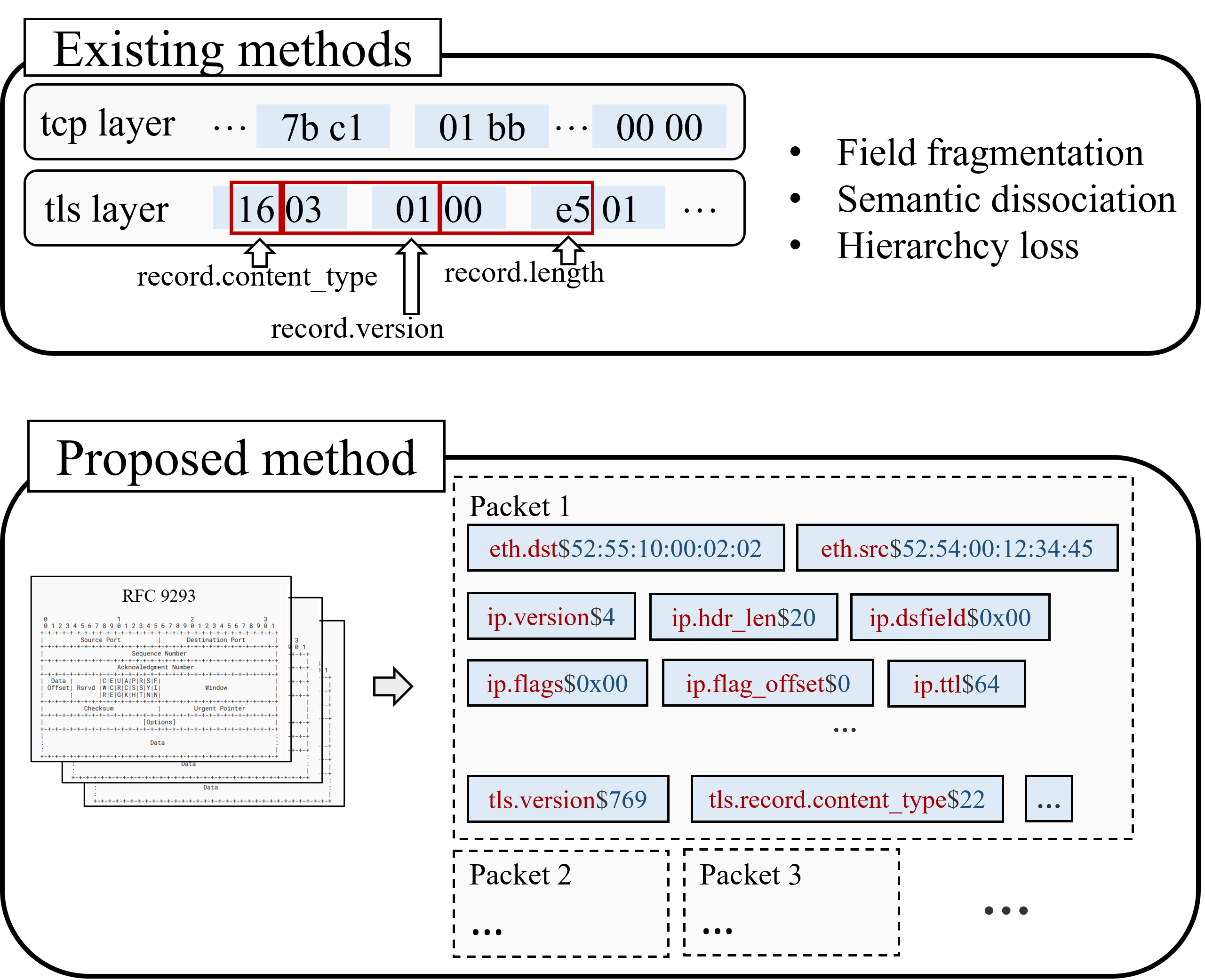}}
	\caption{Tokenization process. Segment traffic data according to protocol specifications, preserving structural and semantic integrity.}
	\label{fig:tok}
\end{figure}

Since this paper focuses specifically on encrypted traffic, we take network packets transmitted under the TLS(Transport Layer Security)  protocol\cite{krawczykOPTLSProtocolTLS2016}, which is the most widely used encryption protocol, as a representative example. TLS is a common protocol used for securing communication over the internet, making it an ideal candidate for demonstrating our proposed method. Specifically, the traffic representation method and the overall tokenization process that we propose in this paper are illustrated in Figure \ref{fig:tok}.

First, protocol fields are determined based on the RFC documentation. Using tools like Scapy\footnote{https://scapy.net/} and Tshark\footnote{https://tshark.dev/}, we extract the relevant structures directly from pcap packets. For fields that are not present in the traffic, corresponding values are set to -1 to ensure the structure remains consistent. Each field and its associated value are concatenated with special symbols, and these field-value pairs are then serialized into pseudo-textual streams, constructing contextualized sequences for the model. This process effectively transforms network traffic packets into semantically enriched, text-like representations, while preserving both the original information and the hierarchical structure of the traffic. By doing so, we enable the model to fully leverage the semantic understanding capabilities of large pre-trained models.

Additionally, while the original GPT-2 vocabulary is primarily designed for natural language understanding, it is not suitable for analyzing network traffic. For instance, the port number 443 might be tokenized as separate digits 4, 4, and 3, losing its intrinsic meaning as a unified service identifier. To address this, we expand the vocabulary to include high-frequency field values and field names specific to network traffic, derived from the datasets we work with. This expansion allows the tokenizer to more accurately segment and encode the processed traffic into embedded vectors with position encodings, ensuring that each field is properly interpreted in the context of network traffic analysis. This approach not only improves the model’s ability to process traffic data but also enhances its performance in generating high-quality representations for traffic generation and classification tasks.

\subsection{Pre-training}

As previously mentioned, unlike conventional tokenization methods, our protocol-based encoding method preserves essential hierarchical relationships by marking the boundaries of protocols and fields using field names. This ensures that the overall structural integrity of network traffic is maintained, allowing the model to retain contextual information throughout the process. By preserving these hierarchical relationships, our approach not only provides a more accurate representation of network traffic but also enables the model to leverage the structure inherent in the data. These domain-specific data representation strategies significantly enhance the model’s performance in network traffic analysis tasks, particularly during the pre-training phase, where the model learns to understand the intricate details of traffic patterns.

For our pre-training process, we use GPT-2 as the base model. This model processes large-scale, unlabeled data through an auto-regressive approach, where each token in the sequence is predicted based on the preceding tokens. During this phase, the model maps each token to a high-dimensional embedding vector, with positional encodings integrated into the vectors to capture the sequential structure of the data. These embedded representations are then passed through a series of stacked Transformer decoder layers, each consisting of a multi-head self-attention mechanism. This mechanism allows the model to simultaneously focus on multiple contextual aspects of the data, which is critical for capturing both local dependencies and long-range relationships within the network traffic.

The residual connections and layer normalization further stabilize the training process and support deeper architectures by ensuring that gradients do not vanish or explode during backpropagation. Additionally, causal masking is applied, which ensures that each prediction made by the model is based only on the preceding tokens, preventing leakage of future information. This layered architectural approach enables the model to effectively learn and represent the complex relationships that exist in network traffic data, making it well-suited for downstream tasks like traffic classification and generation.

Given this architecture, at each time step $t$, the model generates a probability distribution for the next token $x_t$, conditioned on the previous tokens $x_1, x_2, ..., x_{t-1}$. This process is mathematically described as follows:

\begin{equation}
	P(x_t \mid x_1, x_2, \dots, x_{t-1}) = \frac{\exp(o_{t, x_t})}{\sum_{v \in \mathcal{V}} \exp(o_{t, v})}
\end{equation}
where $o_{t,x_t}$ is the logit corresponding to token $x_t$, and $\mathcal{V}$ represents the vocabulary of the model. This equation reflects how the model predicts the likelihood of the next token in the sequence based on the previous context.

For the training process, we employ cross-entropy loss, a widely used method for optimizing the model parameters. The objective function for training is as follows:
\begin{equation}
	\mathcal{L} = - \sum_{t=1}^{T} \log P(x_t \mid x_1, x_2, \dots, x_{t-1})
\end{equation}

This loss function encourages the model to minimize the difference between the predicted token probabilities and the actual token values in the training data, thereby improving its accuracy and performance over time.

Through this architecture and training approach, the model not only acquires the ability to generate contextually appropriate tokens but also gains the flexibility to be fine-tuned for specific downstream tasks, such as traffic classification, thus ensuring its adaptability across a wide range of network traffic analysis applications.

\subsection{Task-specific Application}

The pre-training phase allows the model to learn robust and generalized representations by processing extensive unlabeled network traffic data. This phase helps the model capture the underlying patterns and structures of network traffic without requiring labeled samples.To accommodate different downstream tasks, the model then requires appropriate fine-tuning to specialize in specific applications. This phase involves modifying the model's architecture by adding task-specific layers and adjusting the training objectives to optimize performance for each individual task. This ensures that the model retains the valuable knowledge acquired during pre-training, while also adapting to the unique requirements of the task at hand.

For classification tasks, we introduce a task-specific classification layer on top of the pre-trained model. This layer typically consists of a feed-forward neural network that transforms the model's hidden representations into category-specific logits, representing the model's predictions for each possible class. During supervised training with labeled datasets, the model learns to map input sequences to their corresponding class labels by adjusting its parameters based on the classification error. In addition to standard classification challenges, we also address the issue of class imbalance, which is common in network traffic, particularly in scenarios where some classes (e.g., malicious traffic) are underrepresented. To solve this problem, we introduce traffic generation as a data augmentation strategy. This approach leverages the model's generative capabilities to synthesize additional samples for the underrepresented classes. By generating traffic samples that maintain the statistical and structural characteristics of the target classes, the model can be trained on a more balanced dataset, improving its ability to generalize and develop more robust decision boundaries.

For network traffic generation tasks, we guide the model's generation process by providing a distinct category-specific starting token, which serves as conditional information for the model. This starting token indicates the type of traffic to be generated (e.g., normal or malicious). The model then initiates an auto-regressive generation process, progressively constructing complete network traffic sequences by predicting subsequent tokens. At each time-step, the model uses the previously generated tokens to predict the next most probable token, drawing from its learned understanding of protocol structures and traffic patterns. This auto-regressive mechanism ensures that the generated sequence maintains both coherence and consistency.

\section{EXPERIMENTS}

This section outlines our experimental methodology and provides a detailed analysis of results that validate the model’s effectiveness. We also discuss the current limitations of the model in this section.

\subsection{Settings}

\subsubsection{Datasets}


The model's pre-training, fine-tuning, and validation are conducted using seven public datasets and one dataset composed of self-collected benign traffic, as detailed in Table \ref{tab:datasets2}. The CTU-normal dataset\footnote{https://www.stratosphereips.org/datasets-normal} is constructed from normal traffic data publicly released by Stratosphere, while the MalwareTraffic dataset\footnote{https://www.malware-traffic-analysis.net/} is built based on malicious software traffic data publicly available from Malware-Traffic-Analysis.net. We extract traffic from five classical public datasets (CTU-normal, Datacon, CSTNET-TLS 1.3, ISCXVPN2016, MalwareTraffic) to perform unsupervised pre-training of the model.

\begin{table}[!t]
	\centering
	\caption{Datasets.}
	\label{tab:datasets2}
	\begin{threeparttable}
		\begin{tabular}{l|c|c|c}
			\toprule
			\textbf{Dataset} & \textbf{P} & \textbf{C} & \textbf{G} \\ \midrule
			CTU-normal dataset  & \ding{51} & \ding{55} & \ding{55} \\ \midrule
			Datacon Encrypted Malicious Traffic dataset\cite{datacon2020} & \ding{51} & \ding{51} & \ding{55} \\ \midrule
			ISCXVPN2016\cite{gil2016encrypted} & \ding{51} & \ding{51} & \ding{55} \\ \midrule
			CSTNET-TLS 1.3\cite{linBERTContextualizedDatagram2022} & \ding{51} & \ding{51} & \ding{55} \\ \midrule
			MalwareTraffic dataset & \ding{51} & \ding{55} & \ding{55} \\ \midrule
			EBSNN D1 \cite{xiaoEBSNNExtendedByte2022} & \ding{55} & \ding{51} & \ding{55} \\ \midrule
			CIC IoT dataset 2023\cite{ciciot2023} & \ding{55} & \ding{51} & \ding{51} \\ \midrule
			Self-Collected Benign Traffic & \ding{55} & \ding{55} & \ding{51} \\ \bottomrule
		\end{tabular}
		\begin{tablenotes}
			\footnotesize 
			\item[1] \textbf{P}: used for model pre-training
			\item[2] \textbf{C}: used for traffic classification
			\item[3] \textbf{G}: used for traffic generation and data augmentation
		\end{tablenotes}
	\end{threeparttable}
\end{table}

\subsubsection{Baselines}


We conduct comparative experiments using five different models to thoroughly validate the effectiveness of the proposed model.

\begin{itemize}

	\item FS-Net\cite{liuFSNetFlowSequence2019}: This model converts network traffic into packet length sequences, capturing flow patterns through a multi-layer encoder-decoder architecture that processes sequential data and reconstructs traffic features for classification.
	
	\item DeepPacket\cite{lotfollahiDeepPacketNovel2020}: DeepPacket extracts features from raw traffic using two models: SAE and CNN. Based on the original paper's findings of CNN's superior performance, we implement the CNN approach for our comparative analysis.
	
	\item ET-BERT\cite{linBERTContextualizedDatagram2022}: This method adapts BERT for encrypted traffic analysis by pre-training on large-scale unlabeled data to learn contextual packet representations. Its bidirectional self-attention mechanisms capture complex traffic pattern interdependencies, enabling effective classification across diverse traffic types.
	
	\item YaTC\cite{zhaoAnotherTrafficClassifier2023}: YaTC proposes a Masked Auto-Encoder (MAE) traffic classification framework that converts network traffic into fixed-size matrices for standardized processing. It leverages pre-training mechanisms to extract features and perform accurate classification.
	
	\item NetGPT\cite{mengNetGPTGenerativePretrained2023}: NetGPT employs a GPT-based architecture that processes network traffic as sequential tokens. Through autoregressive training on large-scale datasets, it learns traffic patterns and generates synthetic samples, allowing fine-tuning for various downstream analysis tasks.

\end{itemize}

It should be noted that the first four models lack generative capabilities, and NetGPT only generates specified field types. Thus, we only compare our generative performance with NetGPT.

\subsubsection{Metrics}

In experiments, we employ four commonly used evaluation metrics to comprehensively assess the model's classification performance. Accuracy (AC) quantifies the proportion of correctly classified samples. Precision (PR) measures the ratio of true positives among all positive predictions. Recall (RC) represents the fraction of actual positive samples successfully identified. F1-Score(F1) provides the harmonic mean of precision and recall, offering a balanced measure of overall effectiveness. Collectively, these metrics enable multi-dimensional and objective evaluation of the model's capabilities.

\begin{equation}
	\begin{aligned}
		AC &= \frac{TP + TN}{TP + TN + FP + FN}, \\
		PR &= \frac{TP}{TP + FP}, \\
		RC &= \frac{TP}{TP + FN}, \\
		F1 &= \frac{2 \cdot PR \cdot RC}{PR + RC}
	\end{aligned}
\end{equation}

For the generation task, we evaluate the quality of generated traffic through Jensen-Shannon Divergence (JSD) on key protocol fields, comparing the statistical distributions between real and generated samples. This metric helps validate both the authenticity and accuracy of our generated traffic in terms of protocol-level characteristics.

\subsubsection{Implementation Details}

Our research focuses on encrypted traffic, specifically TLS traffic filtered from datasets using tshark. We exclude categories without TLS traffic to ensure smooth experiment execution. All experiments are implemented using PyTorch and conducted on NVIDIA GeForce RTX 2080 Ti GPU $\times$ 4.

For the classification task, we set the learning rate to 2e-5 and the weight decay to 0.01. The model is fine-tuned for 5 epochs. Specifically, Ethernet and IP addresses are removed to safeguard privacy and improve the reliability of classification results. 

For the generation task, due to the complexity of encrypted traffic and payload invisibility, we focus on processing fundamental fields within the TLS protocol (such as version numbers, record types, etc.) while implementing aggregated generation for payloads. We evaluate our generated traffic's effectiveness on imbalanced classification by creating few-shot scenarios with limited malicious samples. Using test sets with all attacks and balanced benign traffic, we compare models trained with and without generated samples to assess performance improvements under data imbalance.

\subsection{Evaluation on Traffic Generation}

We evaluate our generation model using network traffic from the CIC IoT dataset 2023. We select four representative Web attack types (BrowserHijacking, Backdoor, CommandInjection, and SqlInjection) for the experiments. Erroneous samples identified during the generation process are excluded to ensure data quality. BrowserHijacking represents typical client-side attacks that manipulate user browsing behavior, while Backdoors enable persistent unauthorized system access through malicious code. CommandInjection exploits web application input points to execute harmful commands, and SQLInjection remains among the most prevalent web vulnerabilities. These four attack vectors span both client and server-side security threats in modern web applications. Our selection of these representative attacks enables thorough evaluation of our method's ability to generate diverse attack traffic with practical relevance.

\begin{figure*}[!t]
	\centering
	
	\subfloat[{\fontfamily{ptm}\selectfont BrowserHijacking}]{
		\begin{minipage}[b]{0.48\linewidth}
			\centering
			\begin{minipage}[b]{0.32\linewidth}
				\centering
				\includegraphics[width=\linewidth]{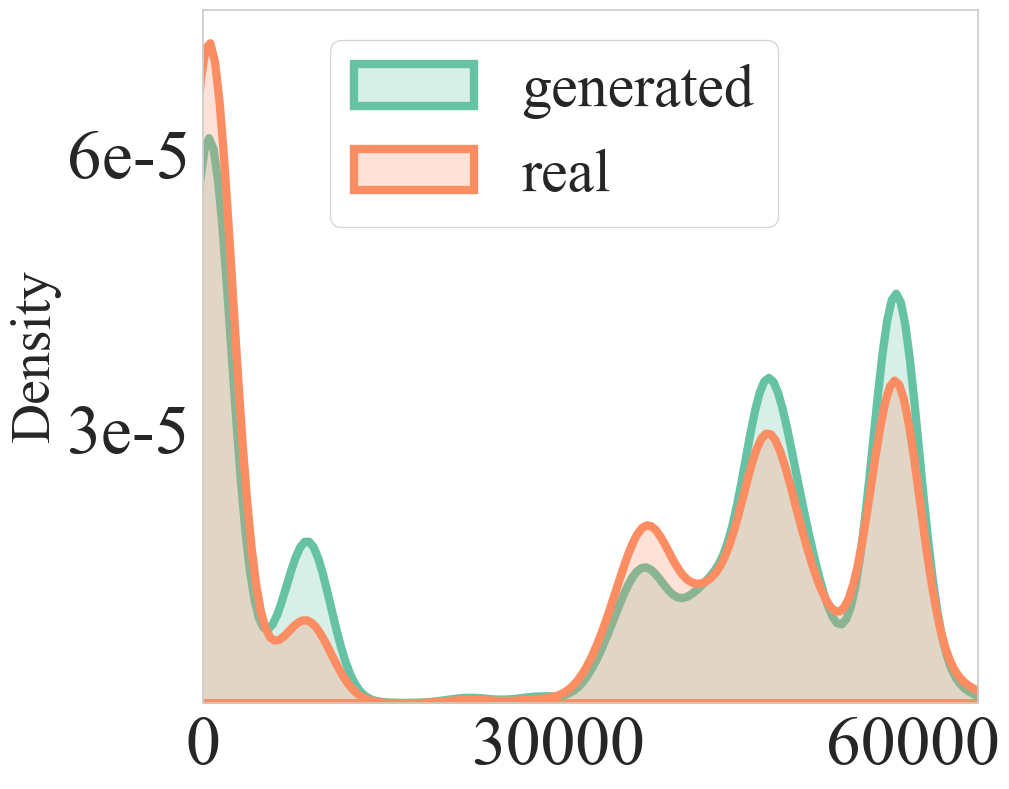}
				\parbox{\linewidth}{\centering sport}
			\end{minipage}
			\hfill
			\begin{minipage}[b]{0.32\linewidth}
				\centering
				\includegraphics[width=\linewidth]{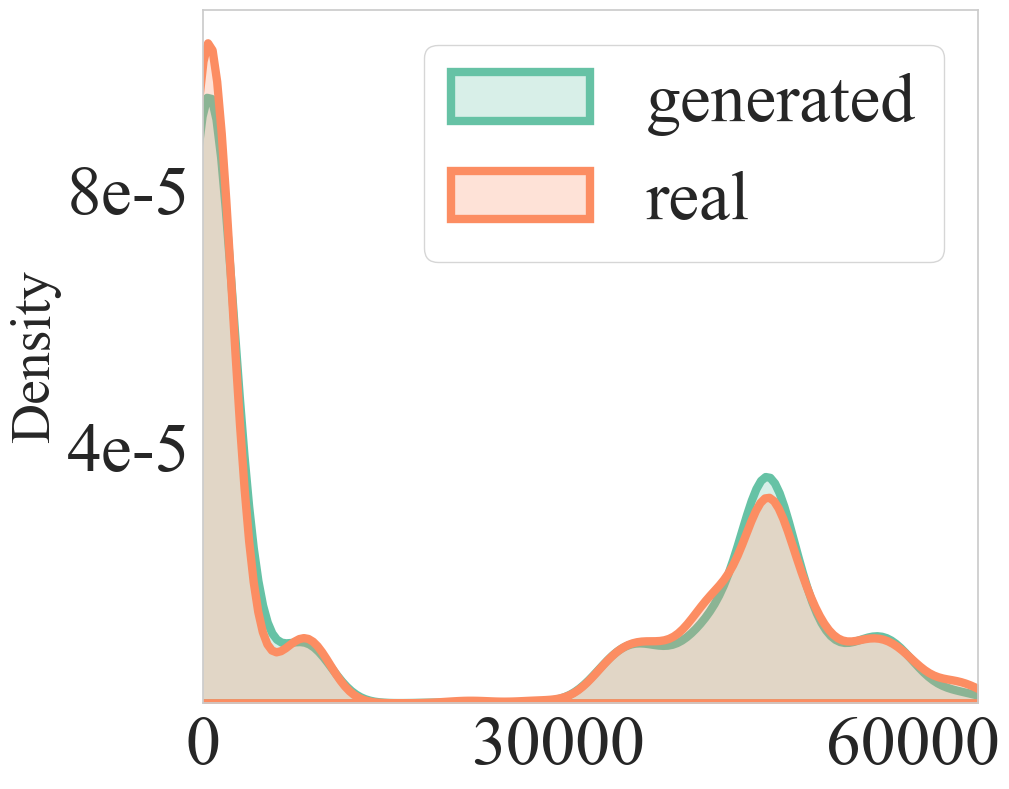}
				\parbox{\linewidth}{\centering dport}
			\end{minipage}
			\hfill
			\begin{minipage}[b]{0.32\linewidth}
				\centering
				\includegraphics[width=\linewidth]{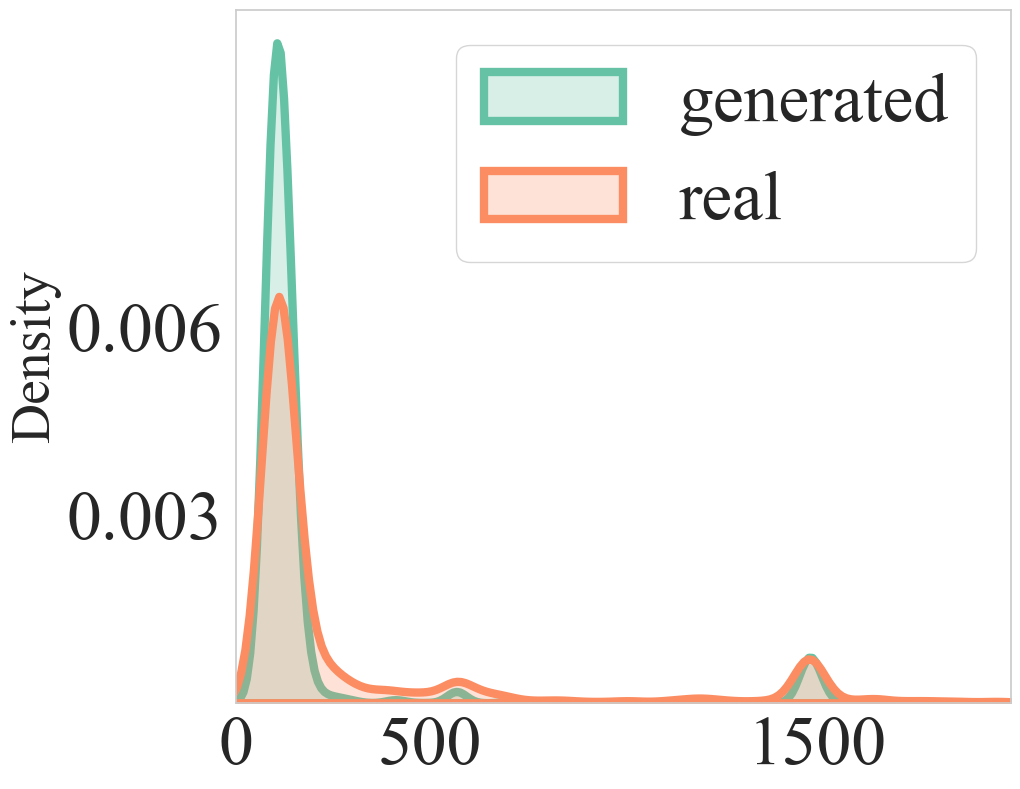}
				\parbox{\linewidth}{\centering len}
			\end{minipage}
		\end{minipage}
	}
	\hfill
	\subfloat[{\fontfamily{ptm}\selectfont Backdoor}]{
		\begin{minipage}[b]{0.48\linewidth}
			\centering
			\begin{minipage}[b]{0.32\linewidth}
				\centering
				\includegraphics[width=\linewidth]{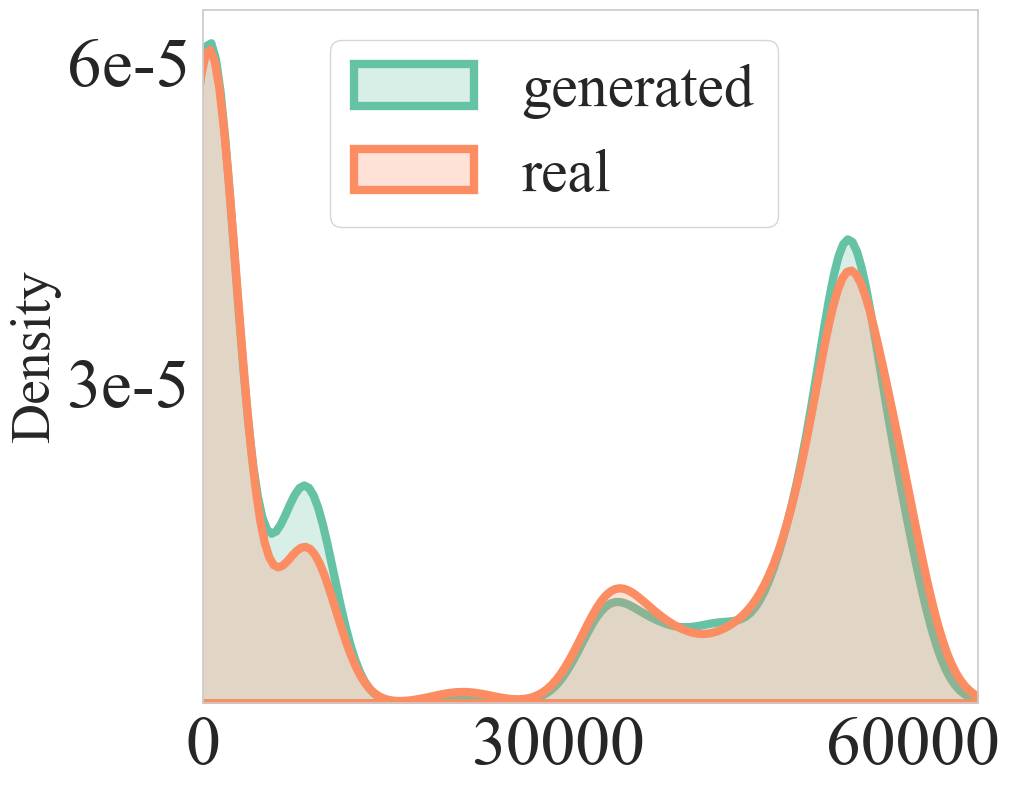}
				\parbox{\linewidth}{\centering sport}
			\end{minipage}
			\hfill
			\begin{minipage}[b]{0.32\linewidth}
				\centering
				\includegraphics[width=\linewidth]{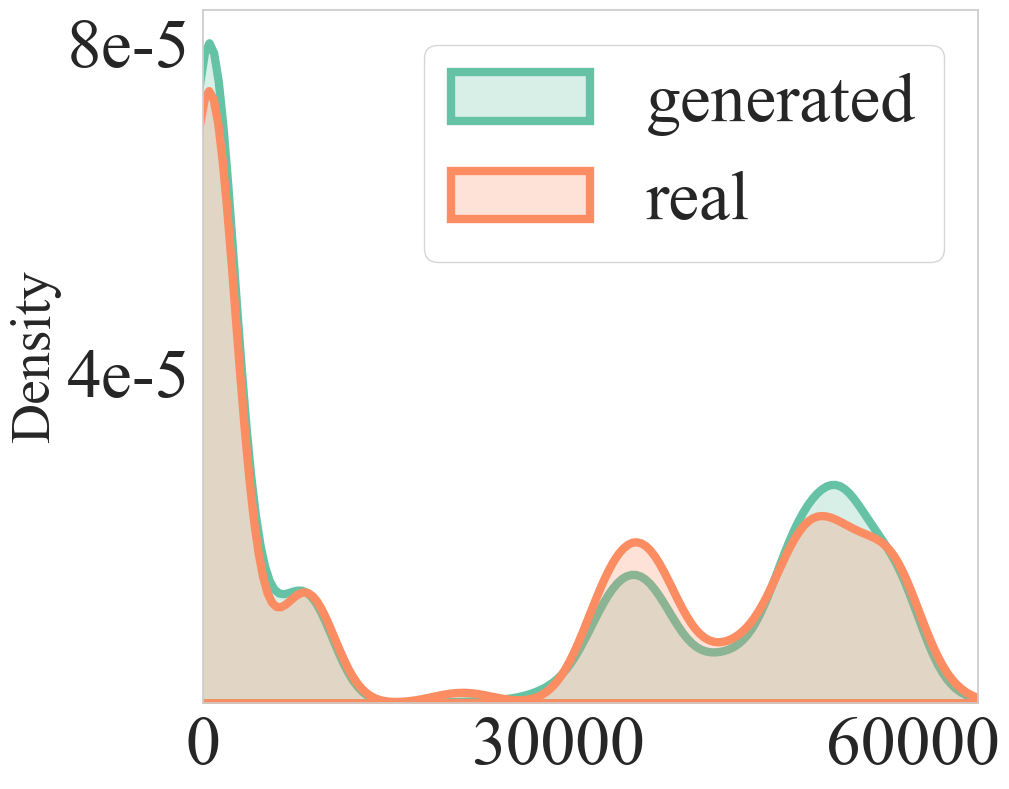}
				\parbox{\linewidth}{\centering dport}
			\end{minipage}
			\hfill
			\begin{minipage}[b]{0.32\linewidth}
				\centering
				\includegraphics[width=\linewidth]{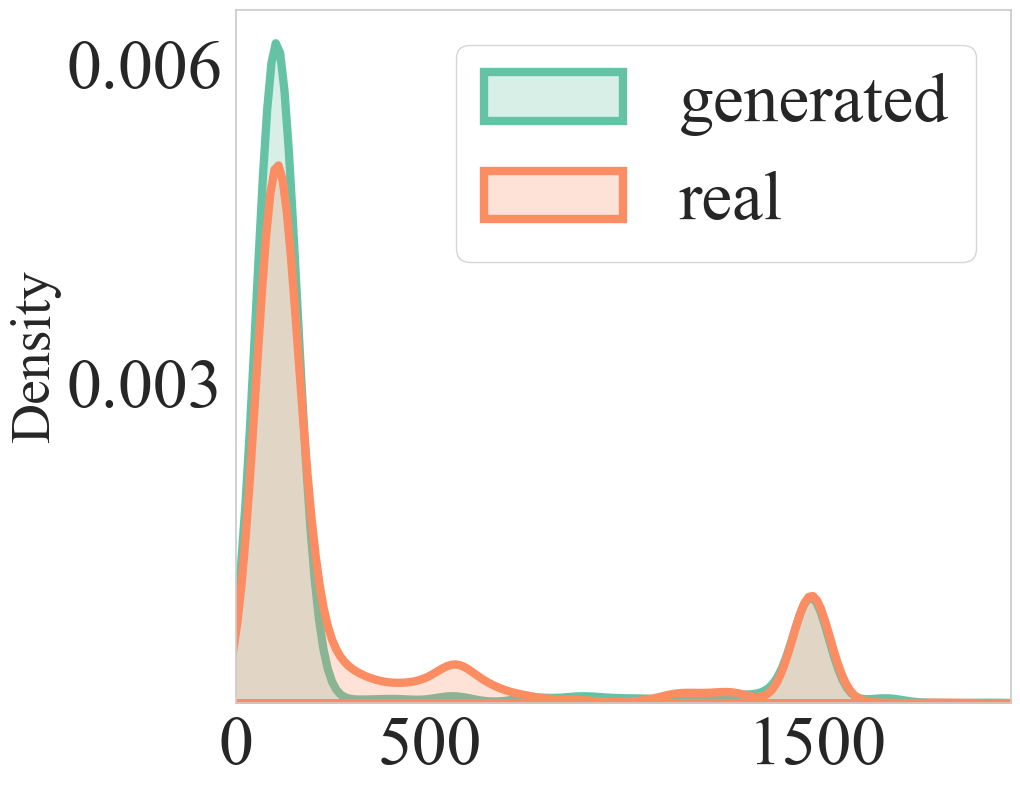}
				\parbox{\linewidth}{\centering len}
			\end{minipage}
		\end{minipage}
	}
	
	\vspace{1em}
	
	\subfloat[{\fontfamily{ptm}\selectfont CommandInjection}]{
		\begin{minipage}[b]{0.48\linewidth}
			\centering
			\begin{minipage}[b]{0.32\linewidth}
				\centering
				\includegraphics[width=\linewidth]{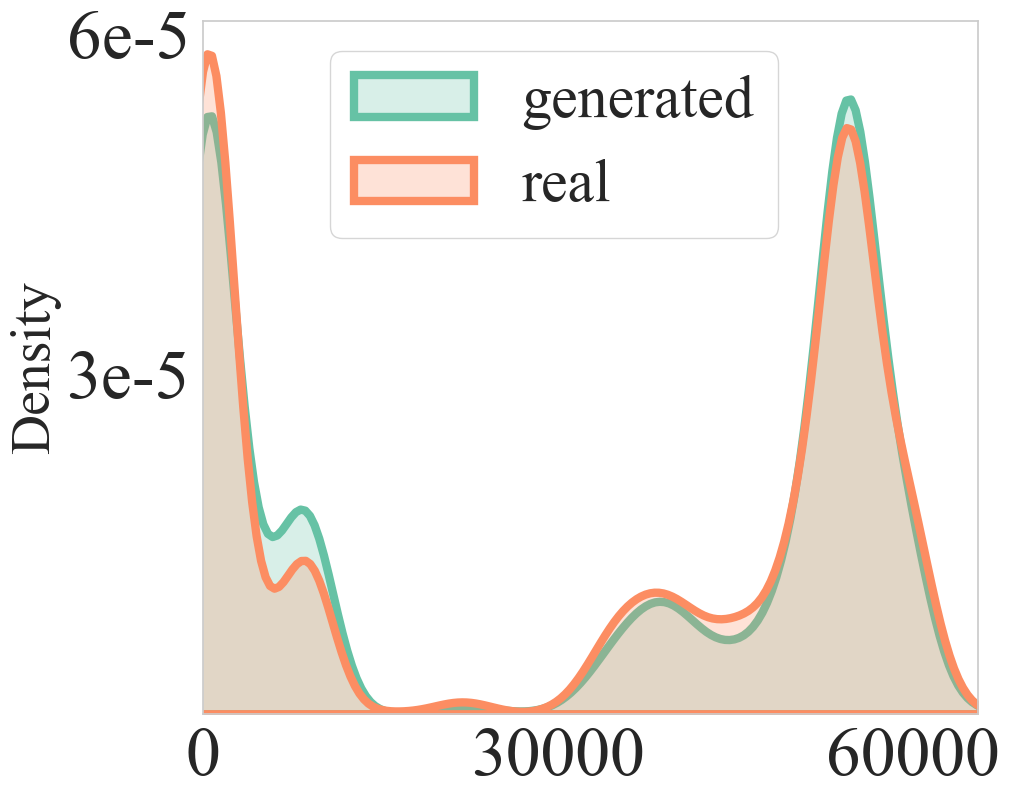}
				\parbox{\linewidth}{\centering sport}
			\end{minipage}
			\hfill
			\begin{minipage}[b]{0.32\linewidth}
				\centering
				\includegraphics[width=\linewidth]{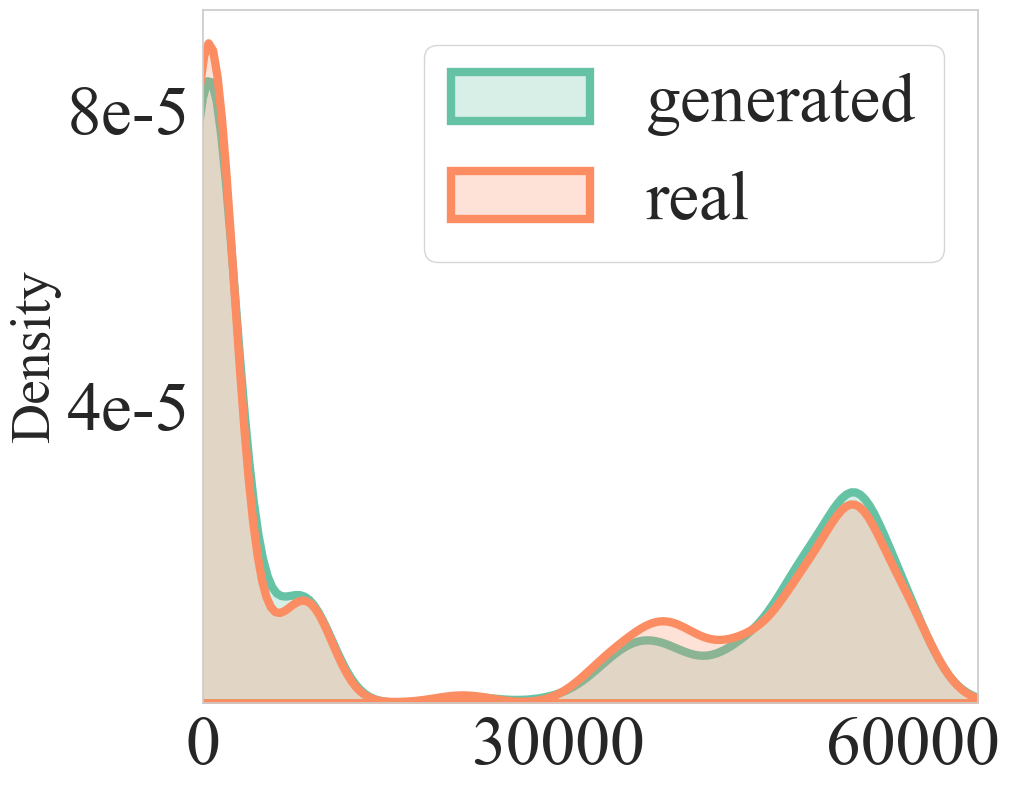}
				\parbox{\linewidth}{\centering dport}
			\end{minipage}
			\hfill
			\begin{minipage}[b]{0.32\linewidth}
				\centering
				\includegraphics[width=\linewidth]{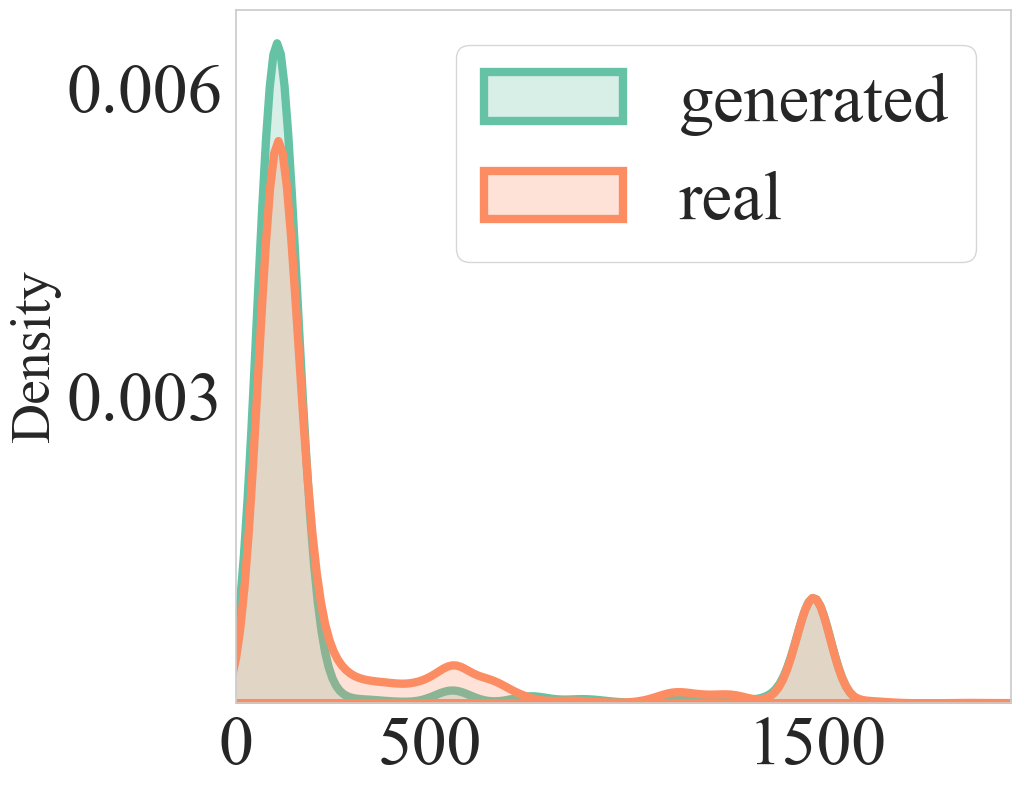}
				\parbox{\linewidth}{\centering len}
			\end{minipage}
		\end{minipage}
	}
	\hfill
	\subfloat[{\fontfamily{ptm}\selectfont SqlInjection}]{
		\begin{minipage}[b]{0.48\linewidth}
			\centering
			\begin{minipage}[b]{0.32\linewidth}
				\centering
				\includegraphics[width=\linewidth]{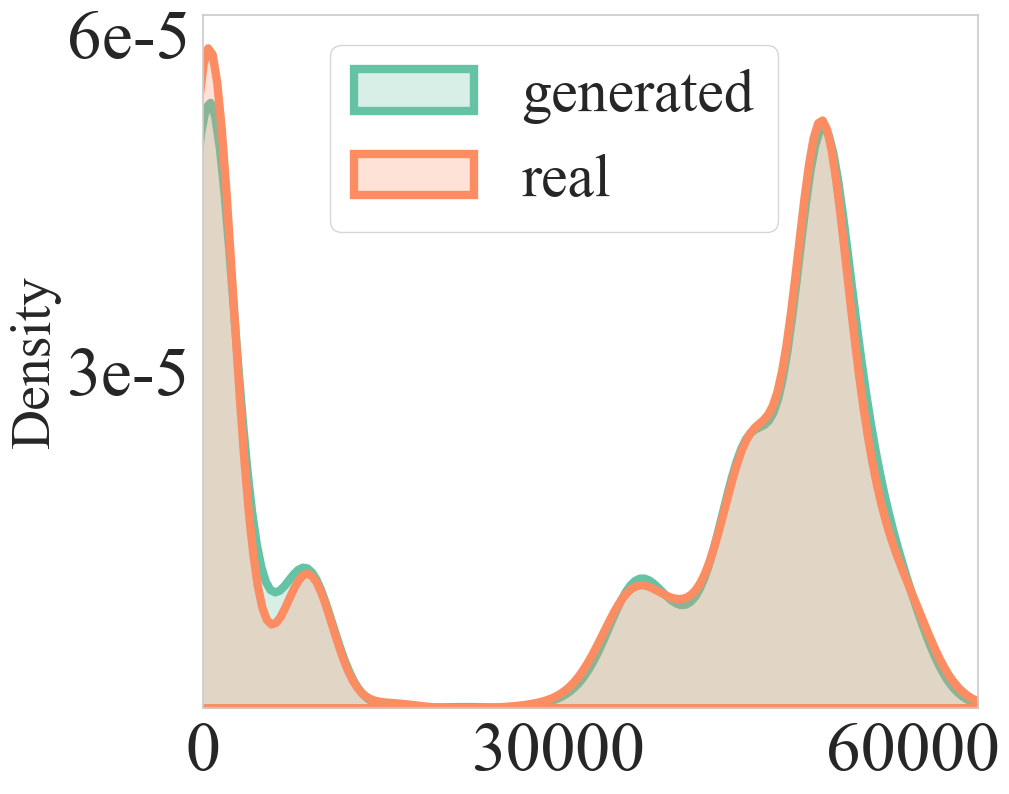}
				\parbox{\linewidth}{\centering sport}
			\end{minipage}
			\hfill
			\begin{minipage}[b]{0.32\linewidth}
				\centering
				\includegraphics[width=\linewidth]{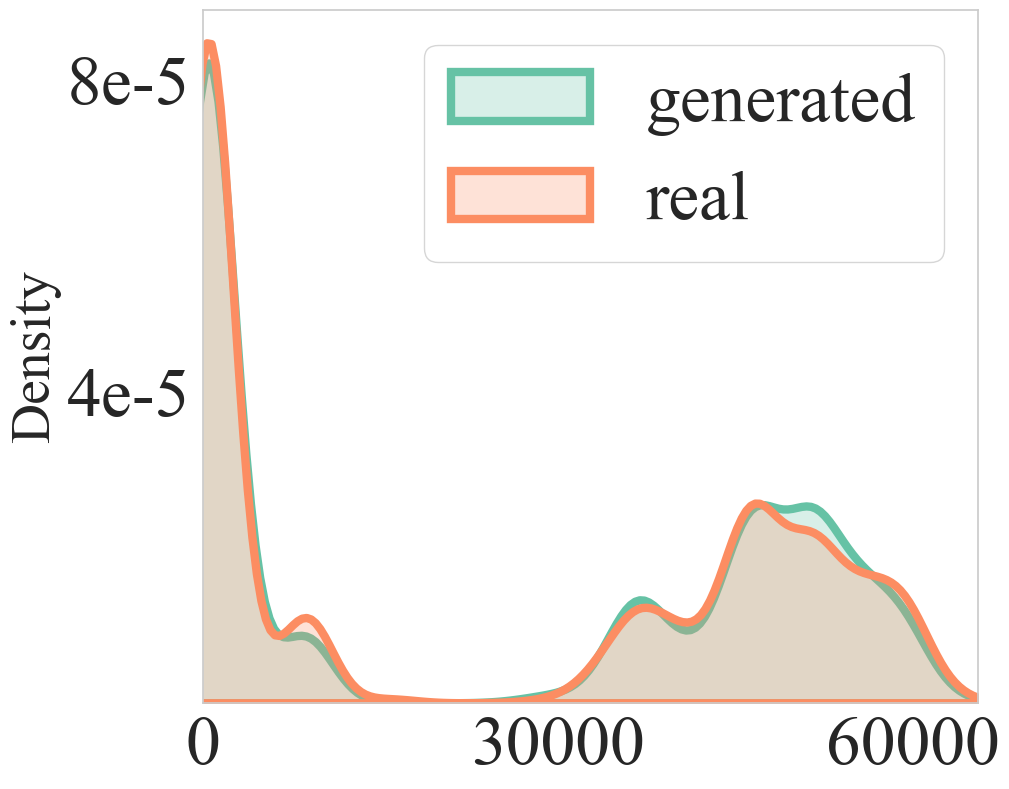}
				\parbox{\linewidth}{\centering dport}
			\end{minipage}
			\hfill
			\begin{minipage}[b]{0.32\linewidth}
				\centering
				\includegraphics[width=\linewidth]{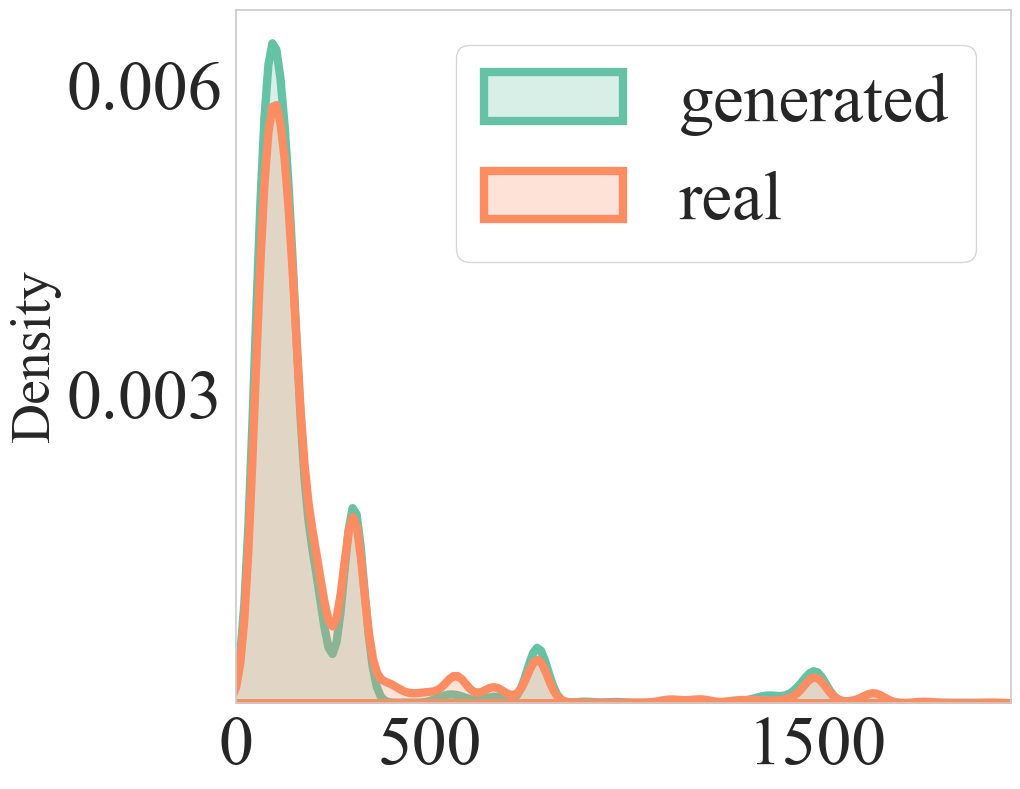}
				\parbox{\linewidth}{\centering len}
			\end{minipage}
		\end{minipage}
	}
	
	\caption{KDE analysis between real and generated traffic. We selecte four types of attacks from the CIC IOT dataset 2023 for our experiments, including BrowserHijacking, Backdoor, CommandInjection, and SqlInjection. We compare the distribution differences between original traffic and generated traffic in terms of source port, destination port, and IP packet length.}
	\label{fig:gen_1}
\end{figure*}

\begin{table}[!t]
	\centering
	\caption{Traffic Generation Performance using JSD.}
	\label{tab:gen_2}
	\begin{tabular}{c|l|c|c}
		\toprule
		\textbf{dataset}                           & \multicolumn{1}{c|}{\textbf{Field}} & \textbf{NetGPT}   & \textbf{GBC} \\ \midrule
		\multirow{3}{*}{BrowserHijacking} & \multicolumn{1}{c|}{sport} & 0.0215 & 0.0221   \\ [0.1em]
		& dport                      & 0.0177 & 0.0165   \\ [0.1em]
		& len                        & 0.0479 & 0.0626   \\ \midrule
		\multirow{3}{*}{Backdoor}         & \multicolumn{1}{c|}{sport} & 0.0360 & 0.0108   \\ [0.1em]
		& dport                      & 0.0214 & 0.0137   \\ [0.1em]
		& len                        & 0.0302 & 0.0353   \\ \midrule
		\multirow{3}{*}{CommandInjection} & \multicolumn{1}{c|}{sport} & 0.0199 & 0.0100   \\ [0.1em]
		& dport                      & 0.0104 & 0.0101   \\ [0.1em]
		& len                        & 0.0474 & 0.0350   \\ \midrule
		\multirow{3}{*}{SqlInjection}     & \multicolumn{1}{c|}{sport} & 0.0522 & 0.0087   \\ [0.1em]
		& dport                      & 0.0631 & 0.0111   \\ [0.1em]
		& len                        & 0.0315 & 0.0311   \\ \bottomrule
	\end{tabular}
\end{table}

To assess the fidelity of generated traffic, we employ Kernel Density Estimate (KDE) analysis, comparing the detailed distribution patterns of critical protocol fields between real and generated samples, as shown in Figure \ref{fig:gen_1}.  From the figure, it can be seen that GBC is capable of generating synthetic traffic that highly fits the distribution of the original traffic.

\begin{figure}[!t]
	\centering
	\subfloat[{\fontfamily{ptm}\selectfont BrowserHijacking}]{
		\includegraphics[width=0.36\columnwidth]{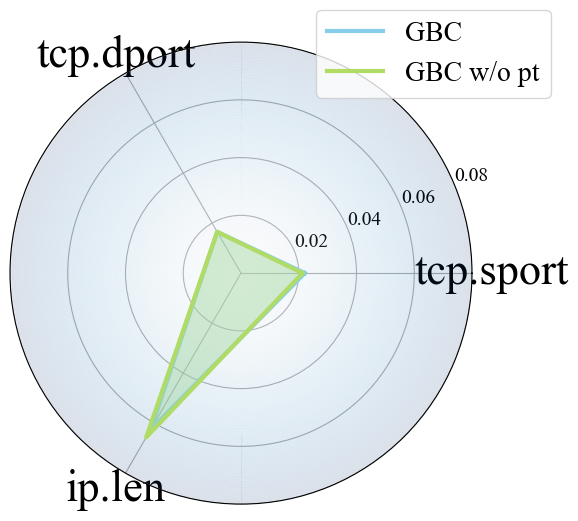} 
	}
	\hspace{0.08\columnwidth}
	\subfloat[{\fontfamily{ptm}\selectfont Backdoor}]{
		\includegraphics[width=0.36\columnwidth]{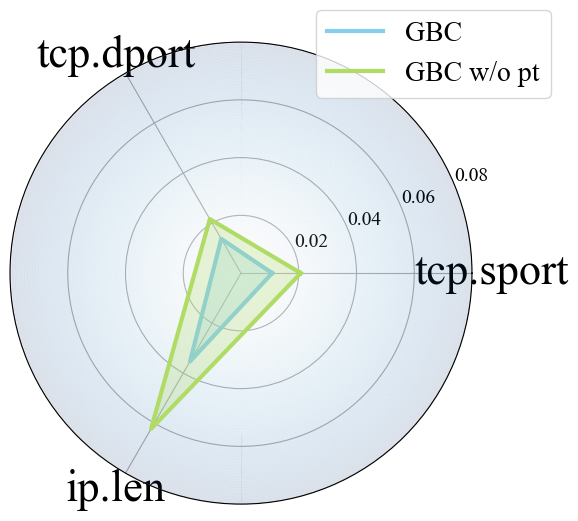} 
	}\\ 
	\subfloat[{\fontfamily{ptm}\selectfont CommandInjection}]{
		\includegraphics[width=0.36\columnwidth]{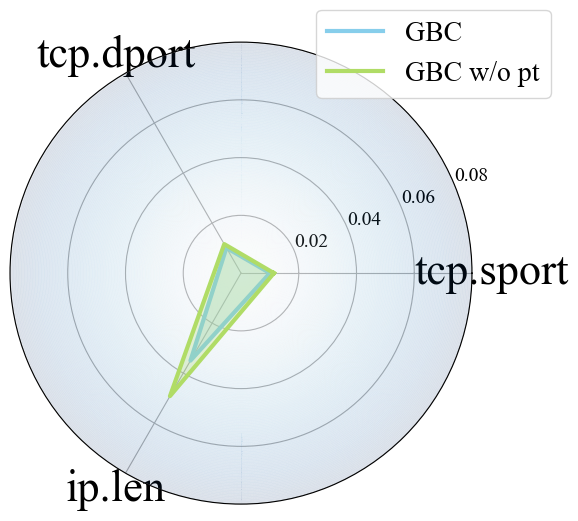} 
	}
	\hspace{0.08\columnwidth}
	\subfloat[{\fontfamily{ptm}\selectfont SqlInjection}]{
		\includegraphics[width=0.36\columnwidth]{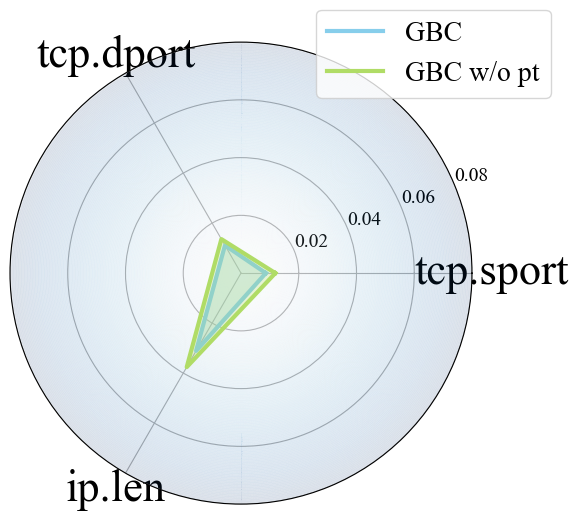} 
	}
	\caption{JSD divergence comparison. GBC w/o pt represents the model with the pre-training step removed. We compare the gap in generation capability between this and the complete model.}
	\label{fig:radio}
\end{figure}

To provide a more intuitive comparison of the distribution similarity between the generated traffic and the original network traffic, we calculate the JSD. This metric quantitatively measures the similarity between two probability distributions, where smaller values indicate that the distributions are very similar, and larger values suggest greater divergence. The detailed JSD results across different protocol fields are presented in Table \ref{tab:gen_2}. It can be observed that compared to NetGPT, our model generates traffic that better matches the original distribution across most categories. While a marginal difference exists in the BrowserHijacking category where our model performs slightly below NetGPT, this gap remains negligible. The primary distinction occurs in packet length distribution, where GBC demonstrates a tendency toward generating shorter packets compared to the more dispersed length distribution observed in original traffic.

Furthermore, to verify the effect of the pre-training process on traffic generation, we compare the performance of the model with the complete architecture versus the model without the pre-training step in the generation task. Figure \ref{fig:radio} shows a comparison of the JSD divergence between the generated traffic and the original traffic for both models. It can be observed that pre-training significantly enhances the model's ability to understand traffic patterns, thereby generating data that better fits the distribution of the original traffic.

It is worth noting that most existing efforts in traffic generation primarily focus on producing partial feature fields (\cite{bikmukhamedovMultiClassNetworkTraffic2021, mengNetGPTGenerativePretrained2023}) or are limited to generating packets under specific and simple protocols (\cite{chengPACGANPacketGeneration2019, kholghPACGPTNovelApproach2023, nukavarapuMirageNetGANbasedFramework2022}). In contrast, the model proposed in this study is capable of generating complete traffic packets of various types. While ensuring the authenticity of the generated samples, it also demonstrates superior generalization, making it applicable to traffic generation tasks across diverse scenarios. 


%
%
%

\newcolumntype{Y}{>{\centering\arraybackslash}X}

\begin{table*}[!t]
	\centering
	\caption{Classification results on Datacon, ISCXVPN2016 and EBSNN D1 datasets.}
	\label{tab:classify_1}
	\begin{tabularx}{\textwidth}{c|YYYY|YYYY|YYYY}
		\toprule
		\textbf{Tasks} & 
		\multicolumn{4}{c|}{\textbf{Malicious Traffic Classification}} & 
		\multicolumn{4}{c|}{\textbf{Encrypted Traffic Classification on VPN}} & 
		\multicolumn{4}{c}{\textbf{Application Traffic Classification}} \\
		\midrule
		Method & AC  & PR  & RC  & F1  & AC  & PR  & RC  & F1  & AC  & PR  & RC  & F1  \\
		\midrule
		FS-Net     & 0.9411 & 0.9465 & 0.8122 & 0.8742 & 0.8085 & 0.6990 & 0.6905 & 0.6946 & 0.8156 & 0.7079 & 0.6657 & 0.6862 \\[0.3em]
		DeepPacket & 0.9049 & 0.9076 & 0.8633 & 0.8848 & 0.7516 & 0.7519 & 0.7516 & 0.7490 & 0.8110 & 0.8185 & 0.8109 & 0.8147 \\[0.3em]
		ET-BERT    & 0.9620 & 0.9620 & 0.9622 & 0.9620 & 0.9840 & 0.9848 & 0.9840 & 0.9842 & \textbf{0.9991} & \textbf{0.9975} & \textbf{0.9864} & \textbf{0.9912} \\[0.3em]
		YaTC       & 0.9485 & 0.9487 & 0.9485 & 0.9473 & 0.9545 & 0.9548 & 0.9514 & 0.9528 & 0.8197 & 0.8035 & 0.7992 & 0.8028 \\[0.3em]
		NetGPT     & 0.8530 & 0.8530 & 0.8564 & 0.8547 & 0.9739 & 0.9524 & 0.9823 & 0.9671 & 0.9667 & 0.8037 & 0.8083 & 0.8060 \\ \midrule
		GBC        & \textbf{0.9776} & \textbf{0.9645} & \textbf{0.9824} & \textbf{0.9734} & \textbf{0.9998} & \textbf{0.9987} & \textbf{0.9982} & \textbf{0.9984} & 0.9982 & 0.9943 & 0.9856 & 0.9900 \\
		\bottomrule
	\end{tabularx}
\end{table*}

\begin{table*}[!t]
	\centering
	\caption{Classification results on CSTNET-TLS 1.3 and CIC IOT 2023 datasets.}
	\label{tab:classify_2}
	\begin{tabularx}{\textwidth}{c|*{4}{>{\centering\arraybackslash}X}|*{4}{>{\centering\arraybackslash}X}}
		\toprule
		\textbf{Tasks} & \multicolumn{4}{c|}{\textbf{Encrypted Application Traffic Classification on TLS 1.3}} & \multicolumn{4}{c}{\textbf{IoT Traffic Classification}} \\
		\midrule
		Method & AC & PR & RC & F1 & AC & PR & RC & F1 \\
		\midrule
		FS-Net     & 0.5988 & 0.6153 & 0.549  & 0.5786 & 0.6445 & 0.5210 & 0.6359 & 0.5726 \\ [0.3em]
		DeepPacket & 0.5380 & 0.5384 & 0.5380 & 0.5377 & 0.7473 & 0.7427 & 0.7473 & 0.7456 \\ [0.3em]
		ET-BERT    & 0.9532 & 0.9535 & 0.9539 & 0.9535 & 0.9801 & 0.9749 & 0.9774 & 0.9761 \\ [0.3em]
		YaTC       & 0.9326 & 0.9331 & 0.9326 & 0.9330 & 0.9368 & 0.9383 & 0.9368 & 0.9383 \\ [0.3em]
		NetGPT     & 0.7011 & 0.6487 & 0.6852 & 0.6665 & 0.8884 & 0.8510 & 0.8570 & 0.8540 \\ 
		\midrule
		GBC        & \textbf{0.9976} & \textbf{0.9975} & \textbf{0.9975} & \textbf{0.9975} & \textbf{0.9805} & \textbf{0.9801} & \textbf{0.9817} & \textbf{0.9810} \\ 
		\bottomrule
	\end{tabularx}
\end{table*}

\subsection{Evaluation on Traffic Classification}

We conduct comprehensive experiments to evaluate the effectiveness of our approach in network traffic classification tasks. Our evaluation consists of two main aspects. First, we assess the model's base performance across five downstream classification tasks on different datasets. Subsequently, we examine how our generated traffic samples can address data imbalance issues through targeted data augmentation experiments. These experiments aim to demonstrate that incorporating strategically generated traffic can effectively improve the model's classification performance, particularly in scenarios where certain attack types have limited real-world training samples.

\subsubsection{Analysis of traffic classification}


The classification performance of proposed model is evaluated through five downstream tasks, each focusing on different aspects of network traffic classification:

\begin{itemize}
	
	\item \textbf{Malicious Traffic Classification:} This task is based on the Datacon Encrypted Malicious Traffic dataset, aiming to achieve binary classification of malicious and normal traffic.
	
	\item \textbf{Encrypted Traffic Classification on VPN:} We utilize the ISCXVPN-2016 dataset, which contains 16 distinct categories, to evaluate the model's effectiveness in identifying traffic transmitted through VPN connections.

	\item \textbf{Application Traffic Classification:} In this study, we utilize the EBSNN D1 dataset, introduced in \cite{xiaoEBSNNExtendedByte2022} and gathered from 29 applications, for application identification.

	\item \textbf{Encrypted Application Traffic Classification on TLS 1.3:} We conduct encrypted traffic application classification experiments on 120 classes based on the CSTNET-TLS 1.3 dataset, as described in \cite{linBERTContextualizedDatagram2022}.

	\item \textbf{IoT Traffic Classification:} This task focuses on encrypted attack classification leveraging the CIC IoT dataset 2023, which includes network traffic data from seven real-world attack categories.

\end{itemize}



The experimental results are presented in Tables \ref{tab:classify_1} and Table \ref{tab:classify_2}. Clearly, the proposed model achieves optimal classification performance on nearly all datasets, demonstrating GBC's superior performance in understanding and analyzing network traffic. 

In the Application Traffic Classification task (EBSNN D1 dataset), the model only achieves the second-best result. Our model is primarily designed for the TLS protocol, while some categories in this dataset contain a large number of data transmission packets using private protocols or tunnels that are difficult to parse. This may have affected the model performance to some extent. Nevertheless, the performance gap between our model and the best-performing ET-BERT in this task is very minimal, which we consider acceptable and could potentially be eliminated through further model optimization. In the other four tasks, our model achieves the optimal classification results. Specifically, in the TLS traffic classification task (CSTNET-TLS 1.3 dataset), the proposed model demonstrates a nearly 5\% improvement on F1 score compared to the second-best ET-BERT. As this dataset comprises TLS traffic across 120 class labels, GBC's excellent performance in this task demonstrates both its applicability in complex scenarios and its effective comprehension of the TLS protocol.

\begin{figure}[!t]
	\centering
	\includegraphics[width=0.85\columnwidth]{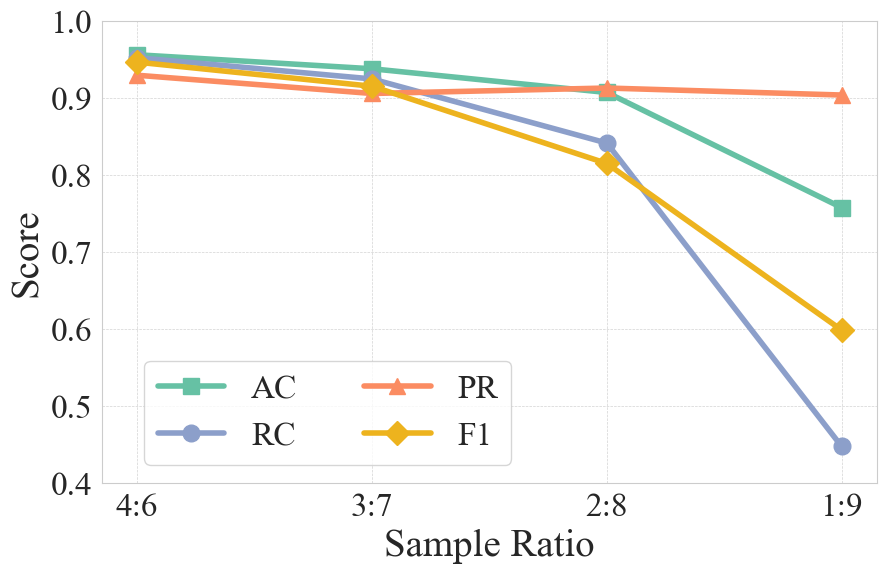}
	\caption{Changes in classification performance under different positive-to-negative sample ratios. From left to right, the proportion of positive samples gradually decreases.}
	\label{fig:imba}
\end{figure}

Moreover, pre-trained models (including ET-BERT, YaTC, NetGPT, and the proposed model) significantly outperform traditional models across all evaluation tasks. In the latter four tasks, pre-trained models demonstrate performance improvements of over 10\% compared to non-pre-trained models. Since the Malicious Traffic Classification task(Datacon Encrypted Malicious Traffic dataset) is a binary classification task determining whether network traffic is malicious, traditional models also achieve good results, making performance differences relatively less pronounced. Nevertheless, pre-trained models still attain the optimal results. These findings underscore the critical role of pre-training in the field of traffic analysis. Such performance gaps suggest that pre-training enables models to learn more robust and transferable features from network traffic data.

Notably, the CIC IoT dataset 2023 and EBSNN D1 datasets are not included in the model's initial pre-training process. Despite this, the proposed model still achieves outstanding performance in the corresponding tasks, demonstrating its exceptional generalization ability in traffic comprehension. The robust performance on datasets not used in pre-training indicates that our model has successfully learned general traffic patterns and characteristics. These results collectively validate the effectiveness of our model design, suggesting promising potential for practical applications in diverse network environments.

\begin{figure}[!t]
	\centering
	\includegraphics[width=0.9\columnwidth]{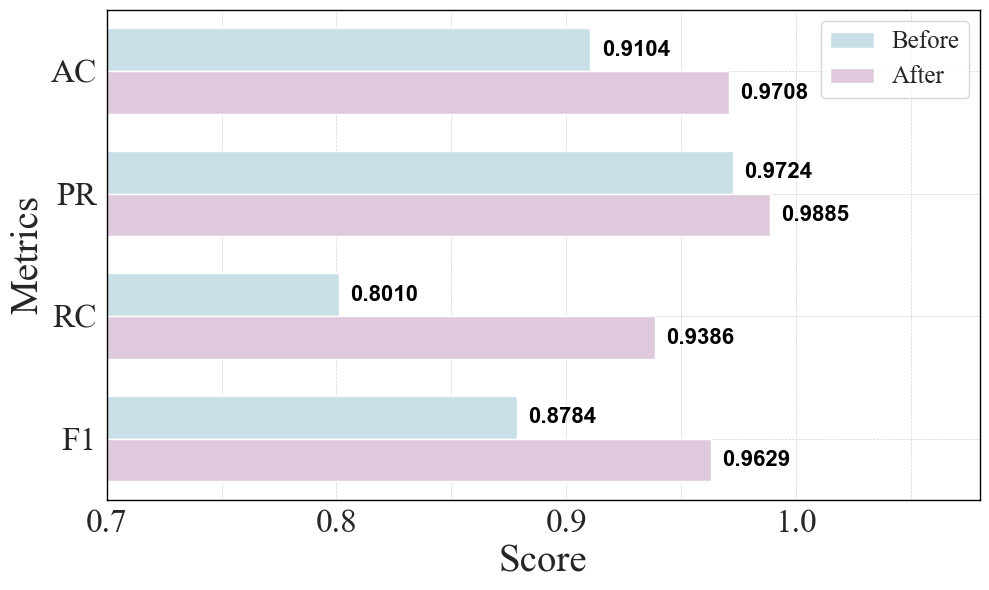} 
	\caption{Comparison of classification performance before and after augmentation. The original ratio of malicious samples to benign samples is 1:5, which achieves balance after expansion with generated traffic.}
	\label{fig:aug}
\end{figure}

\subsubsection{Data Augmentation}

Class imbalance remains a critical and persistent challenge in network traffic classification, where certain attack types often have fewer samples than benign traffic. This imbalance typically leads to poor detection performance for minority classes. To systematically evaluate this impact, we construct an experimental dataset using the CIC IoT dataset 2023. As shown in Figure \ref{fig:imba}, the model's classification performance varies significantly under different positive-to-negative sample ratios, highlighting the detrimental effect of class imbalance on model accuracy. The results clearly demonstrate that as the sample ratio becomes more imbalanced, all performance metrics of the model substantially decrease.

To address this limitation, we propose a traffic generation approach specifically designed to augment minority classes. To evaluate our proposed method, we conduct data augmentation experiments using a subset of the CIC IoT dataset 2023, where we  combine randomly selected Web attack samples with benign samples from both the original dataset and our self-collected normal traffic. These experiments simulate real-world data scarcity scenarios and demonstrate the effectiveness of our generated traffic in improving overall classification performance across multiple evaluation metrics.

\textbf{Data Augmentation:} Specifically, we construct the initial training dataset with a subset of attack samples and benign traffic, maintaining an imbalanced ratio of approximately 1:5 between attack and benign samples. We then generate additional attack samples based on these data to balance the class distribution. Note that the amount of real traffic remains unchanged before and after augmentation, but the augmented dataset contains a larger number of model-generated malicious samples compared to the pre-augmentation dataset. For evaluation, we create a balanced test dataset comprising attack samples and an equal amount of benign traffic. Figure \ref{fig:aug} demonstrates how our synthetic samples improve the model's detection capability compared to training with the imbalanced dataset.

\begin{figure*}[!htbp]
	\centering
	
	\subfloat[{\fontfamily{ptm}\selectfont Accuracy}]{%
		\includegraphics[width=0.23\textwidth]{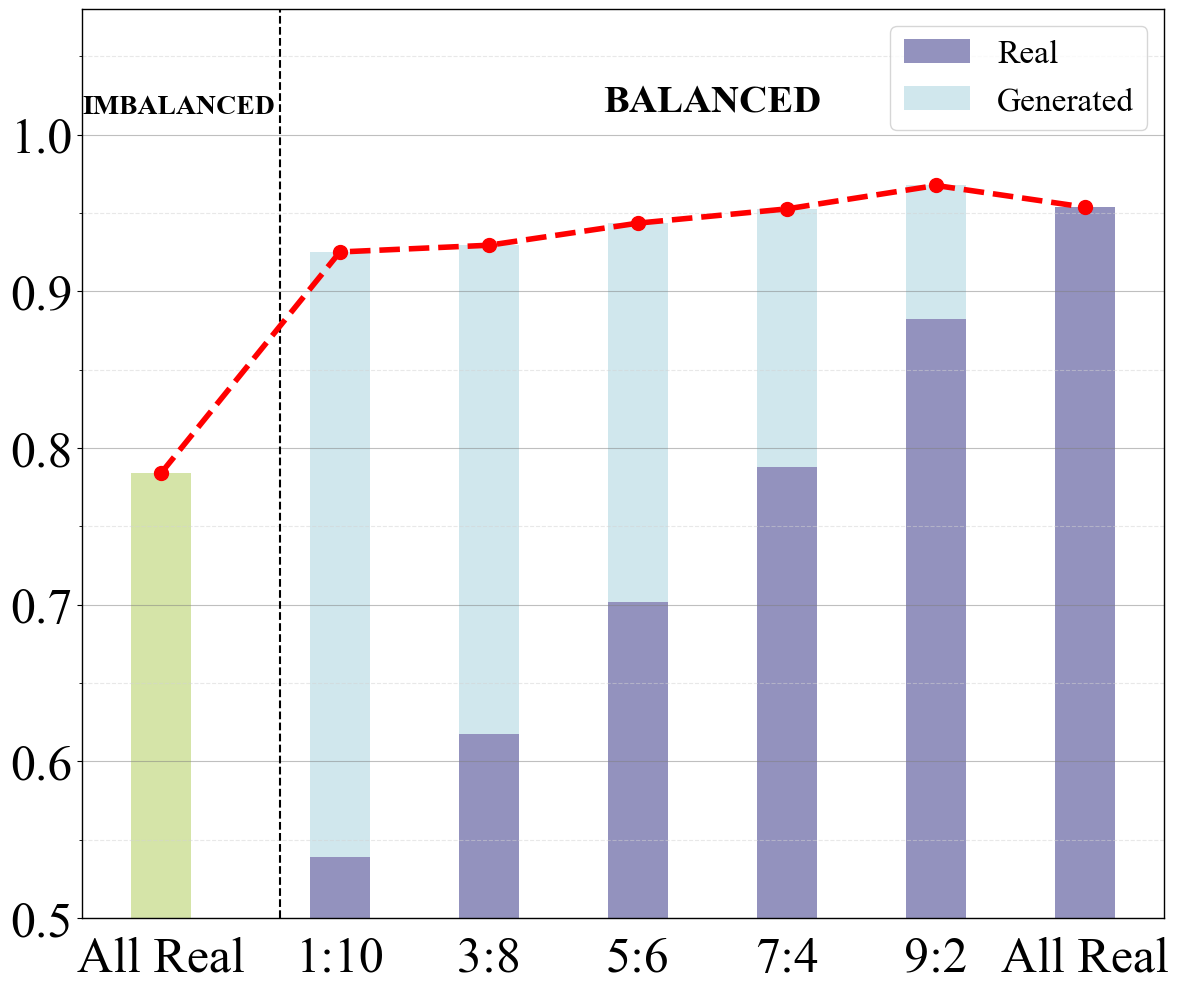}
	}
	\hfill
	\subfloat[{\fontfamily{ptm}\selectfont Precision}]{%
		\includegraphics[width=0.23\textwidth]{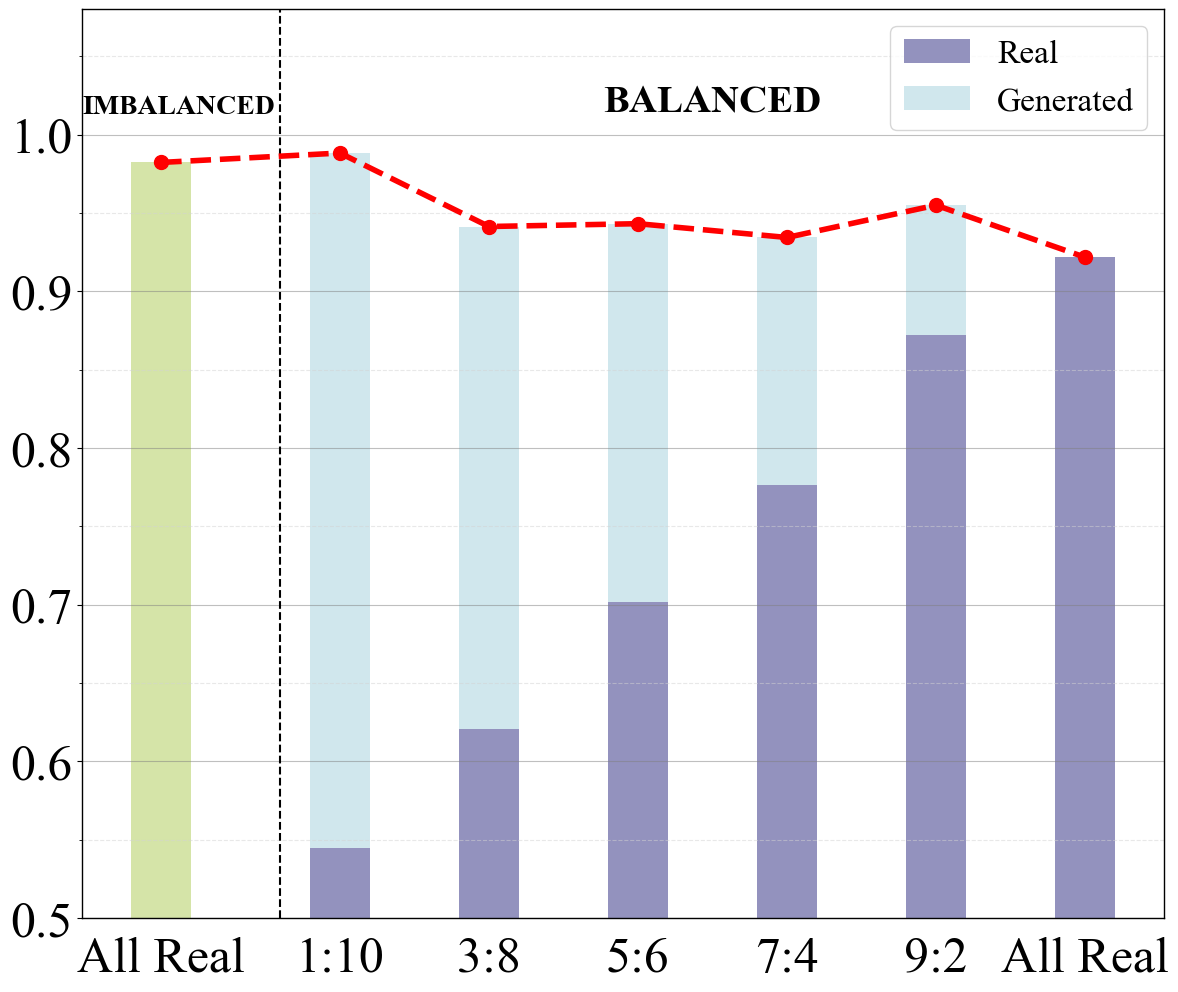}
	}
	\hfill
	\subfloat[{\fontfamily{ptm}\selectfont Recall}]{%
		\includegraphics[width=0.23\textwidth]{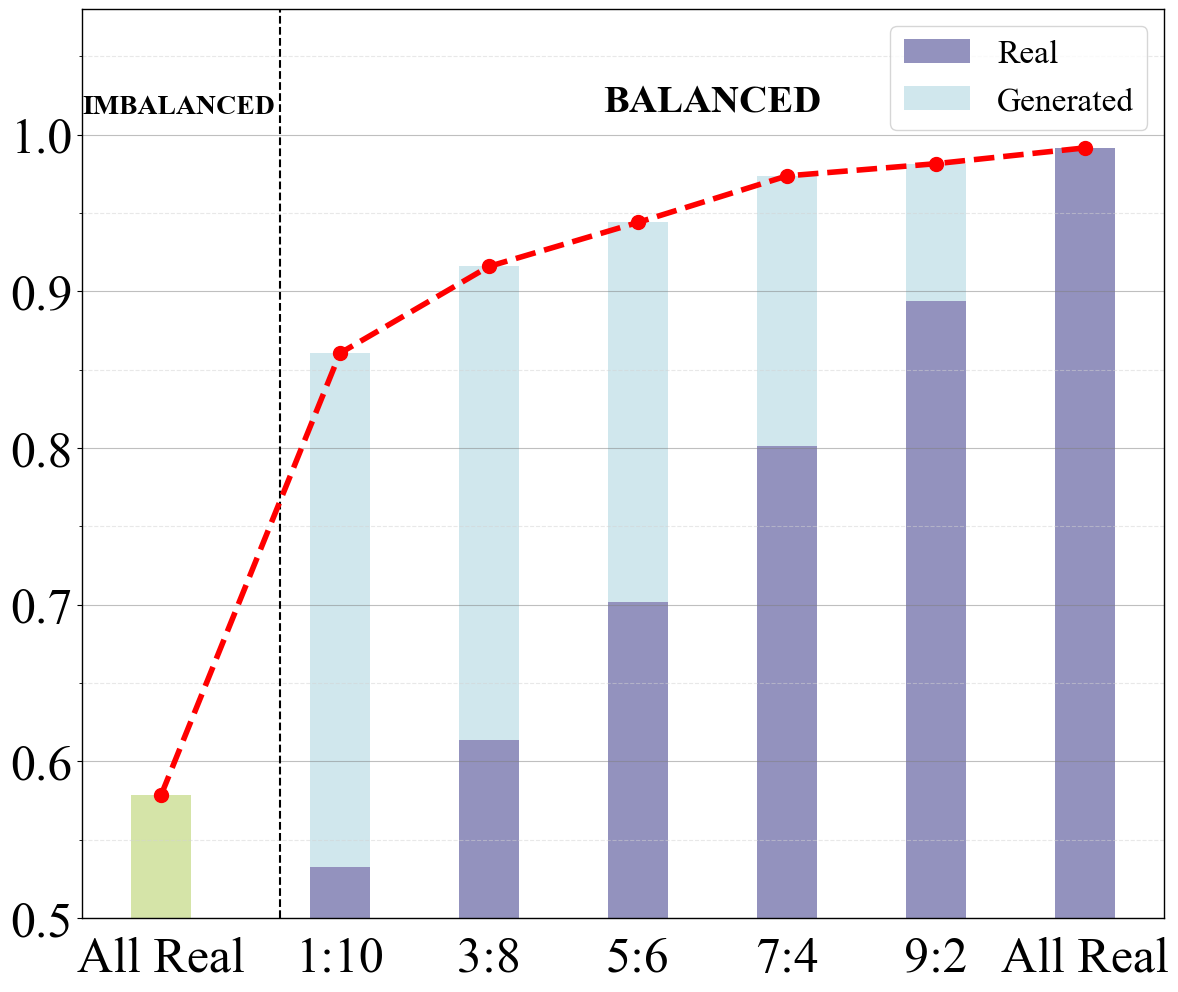}
	}
	\hfill
	\subfloat[{\fontfamily{ptm}\selectfont F1}]{%
		\includegraphics[width=0.23\textwidth]{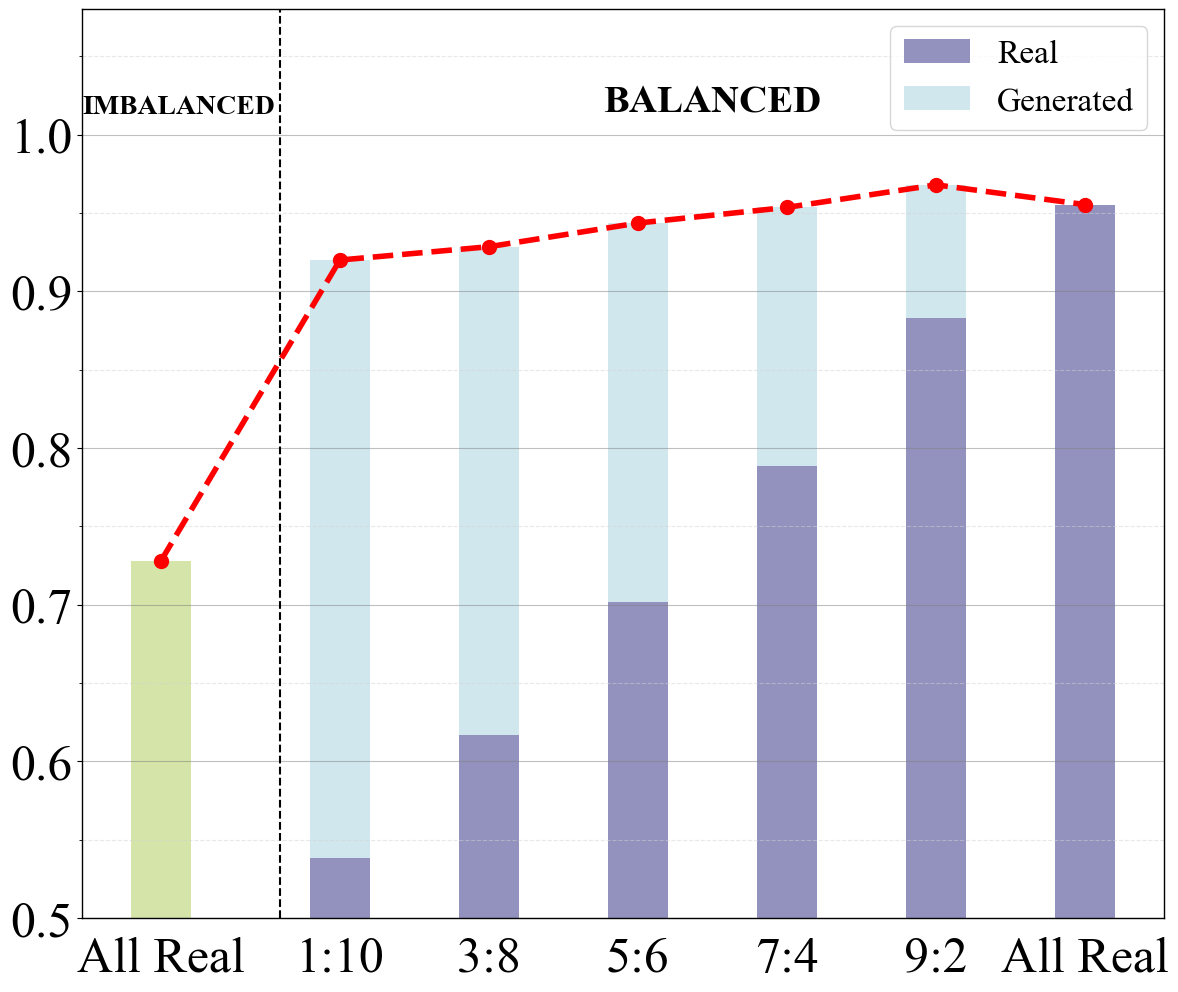}
	}
	
	\caption{Comparison of model performance under different sample proportions. To make the results more intuitive, in this set of experiments, we intensify the degree of imbalance, with the original ratio of malicious traffic to benign traffic being 1:10. In the Balanced part, sections of different colors in the bars represent the proportions of real traffic and generated traffic within malicious samples. From left to right, the proportion of real traffic gradually increases, until the malicious part is entirely composed of real traffic in the final bar.}
	\label{fig:enhance4}
\end{figure*}

\begin{table}[!t]
	\centering
	\caption{Comparison of model performance under different sample proportions.}
	\begin{tabular}{c|c|c|c}
		\toprule
		\textbf{Positive-to-Negative Sample}                      & \textbf{Real : Generated} &  \textbf{TPR}    & \textbf{AUC}    \\ \midrule
		\multicolumn{2}{c|}{\makecell[c]{IMBALANCED \\ malicious:benign=1:10}}  & 0.6252 & 0.9729 \\ \midrule
		\multirow{6}{*}{BALANCED} & 1:10                            & 0.8604 & 0.9905 \\  [0.3em]
		& 3:8                             & 0.9158 & 0.9817 \\ [0.3em]
		& 5:6                              & 0.9439 & 0.9860 \\ [0.3em]
		& 7:4                            & 0.9736 & 0.9891 \\ [0.3em]
		& 9:2                             & 0.9814 & 0.9938 \\ [0.3em]
		& All real                        & 0.9915 & 0.9924 \\ \bottomrule
	\end{tabular}
	\label{tab:enhance4}
\end{table}

Through data augmentation, the F1 score of the classification model improves by 9\%, indicating that our generated samples effectively preserve the critical features of the original traffic. The improved performance mainly comes from the more balanced sample distribution created by adding generated traffic, allowing the model to better learn features from more diverse training samples. This also corresponds with the conclusion in Figure \ref{fig:imba}, which shows that a more balanced sample ratio enables the model to achieve better classification results.

\textbf{Effect of Generated Traffic Ratio:} To further validate the quality of our generated traffic, we conduct an experiment in which real and synthetic traffic are mixed in varying proportions for model training. Figure \ref{fig:enhance4} and table \ref{tab:enhance4} illustrate the results of the experiment. The leftmost bar represents the baseline classification performance under imbalanced conditions (1:10 ratio of malicious to benign samples) using only real traffic. To the right of the dashed line, we present improved classification results after addressing this imbalance by supplementing with our generated samples to achieve a balanced 1:1 ratio. These bars illustrate performance across varying proportions of real and generated data, with the fraction of real samples increasing progressively from left to right.

\begin{figure}[!t]
	\centering
	\subfloat[{\fontfamily{ptm}\selectfont IMBALANCED}]{
		\includegraphics[width=0.31\columnwidth]{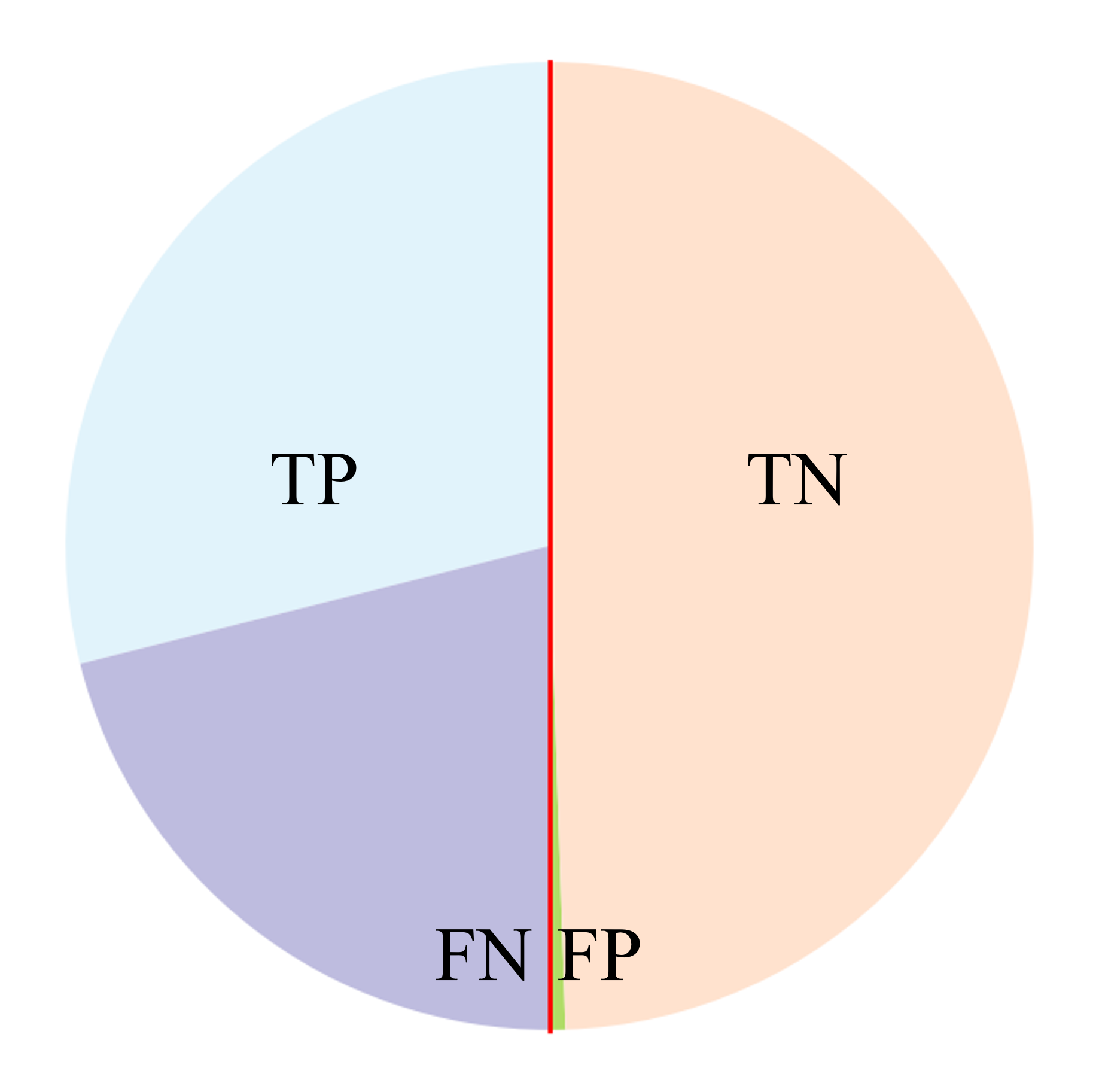} 
	}
	\subfloat[{\fontfamily{ptm}\selectfont BALANCED}]{
		\includegraphics[width=0.31\columnwidth]{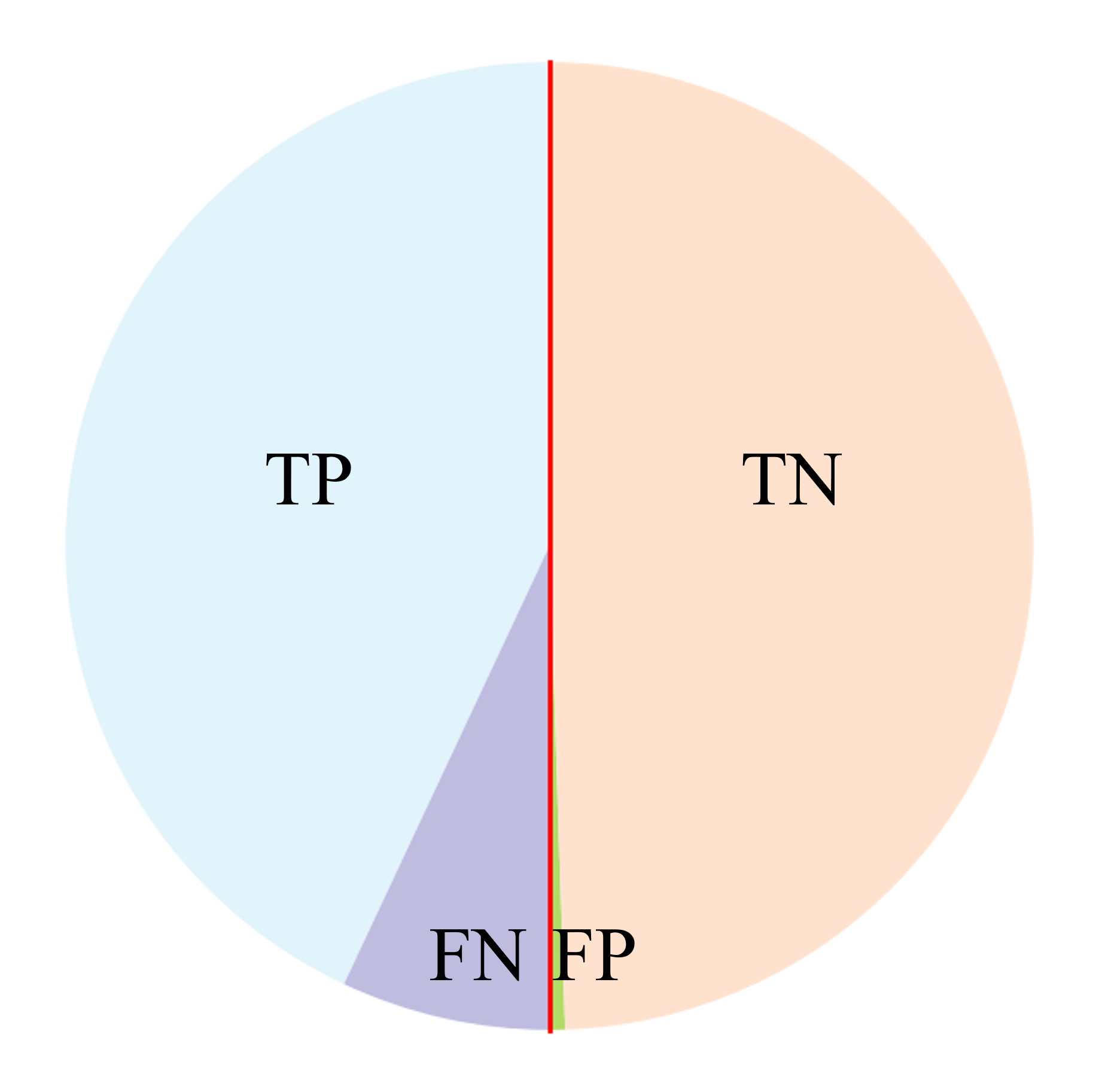} 
	}
	\subfloat[{\fontfamily{ptm}\selectfont BALANCED}]{
		\includegraphics[width=0.31\columnwidth]{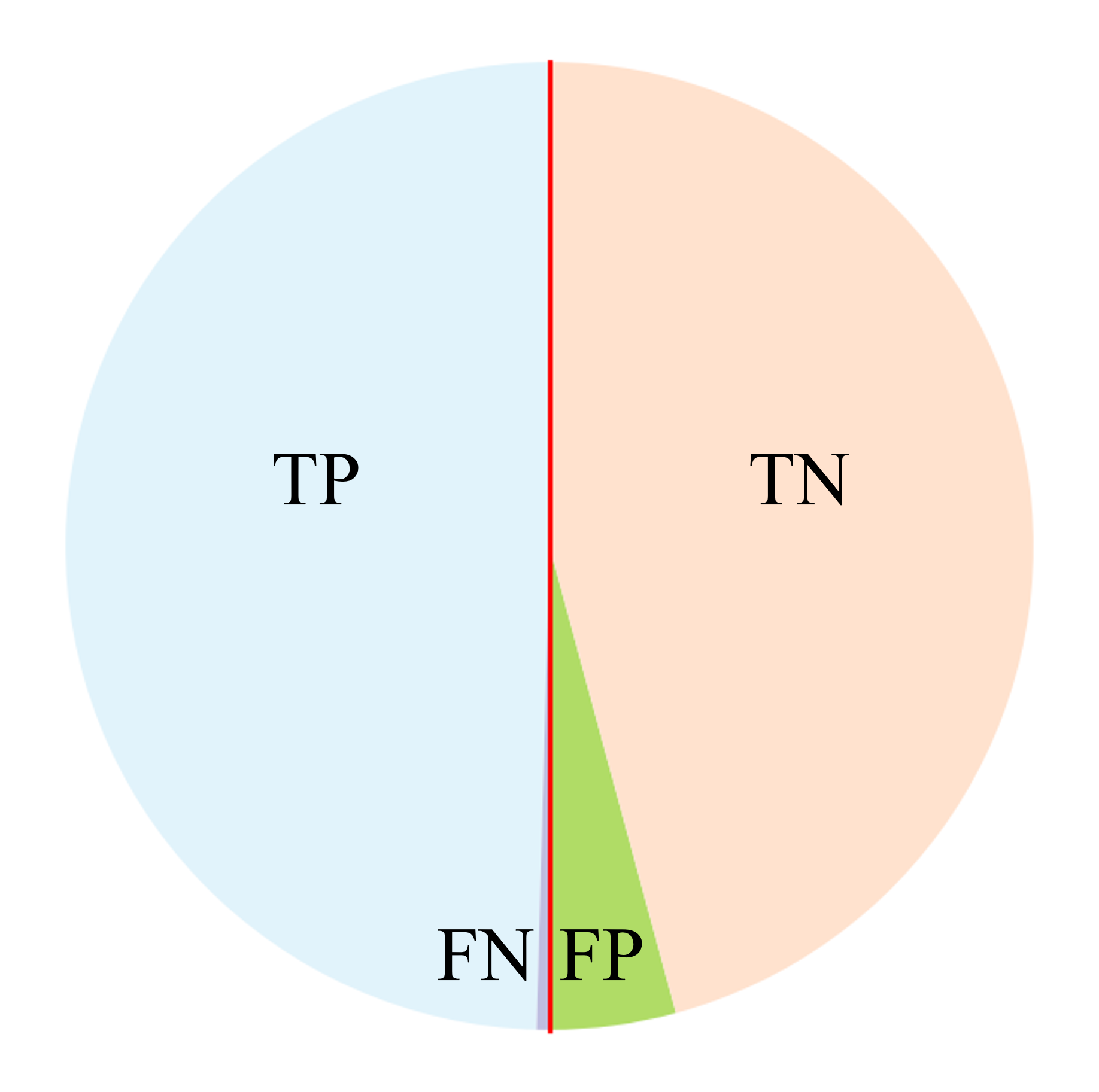} 
	}
	\caption{(a) shows the prediction results of the model trained on the imbalanced dataset. Since the model tends to predict samples as the majority class, FP is low and precision is high. (b) displays the prediction results of the model trained on a balanced dataset after augmentation with generated traffic, where FN is significantly reduced and the model can correctly identify malicious samples. (c) shows the prediction results of the model after being trained on a balanced dataset composed entirely of real traffic samples.}
	\label{fig:circle}
\end{figure}

It is notable that although the number of real malicious samples remains consistent between the imbalanced condition and the balanced condition (with a 1:10 ratio of real to generated samples), the model’s performance improves significantly. This improvement demonstrates that the generated traffic effectively simulates the characteristics of real malicious traffic, augmenting the training dataset.  When compared to a balanced dataset composed solely of real malicious samples, there is an inevitable performance decline as the proportion of generated traffic increases. However, given the practical challenges of acquiring real traffic and the convenience of generating synthetic data, we consider this minor loss in performance entirely acceptable.

In addition, from the figure we can observe that as the sample ratio becomes balanced, Precision instead exhibits a downward trend. We believe this occurs because when the sample ratio is imbalanced, normal traffic samples (negative) constitute the majority. Figure \ref{fig:circle} illustrates the model's prediction results under imbalanced sample conditions and after balancing with generated traffic. After training with the imbalanced dataset, the model tends to classify most samples as normal, thereby resulting in lower FP and higher Precision. With the dataset augmented by generated traffic, the sample distribution tends to be balanced, enabling the model to better learn the features of malicious traffic and thereby achieve effective differentiation.

In summary, the results presented above validate the effectiveness of our traffic generation approach in addressing class imbalance issues, particularly in strengthening the model's capability to detect minority attack classes. This improvement demonstrates the practical applicability of our method in network traffic analysis, where data imbalance is a persistent challenge.

\subsection{Discussion}

Despite GBC's impressive capabilities in both traffic generation and classification tasks, the model does have several limitations to consider. 

\textbf{Resource Consumption:} As a pre-trained model, it inevitably relies on GPU resources and requires more computational overhead compared to non-pre-trained models, which presents potential deployment challenges in resource-constrained environments. 

\textbf{Model Capacity:} Considering the hardware limitations and the computational resource requirements, the model has a relatively small parameter count, which limits its performance in certain complex scenarios. This suggests that there is room for improvement in both the architecture and performance of the model.  

\textbf{Generalizability:} In this study, we only investigate GBC's performance in TLS traffic scenarios, leaving exploration of other encryption protocols for future work. Furthermore, our handling of the TLS protocol is relatively coarse-grained, with insufficient support for some complex protocol fields. Given that TLS is currently the most widely used encryption protocol, we believe that our experimental results based on TLS are representative. Additionally, our protocol-aware tokenization scheme can be easily modified to support additional protocol types.

\section{CONCLUSION AND FUTURE WORKS}

In this paper, we propose a traffic comprehension model based on a generative pre-training architecture, which supports two core capabilities: synthetic traffic generation and encrypted traffic classification. These capabilities enable the model to tackle both the creation of network traffic samples and the effective classification of encrypted traffic. To achieve this, we introduce a novel network traffic representation method specifically designed for pre-training tasks. This method improves the model's ability to understand the underlying semantics of network traffic and capture essential features, such as protocol structures and traffic patterns, which are crucial for accurate traffic analysis.

Our evaluations across five diverse datasets demonstrate that the model performs exceptionally well in traffic classification. More importantly, the model's traffic generation capability proves to be effective for data augmentation, producing synthetic traffic samples that closely resemble real-world network flows. These synthetic samples effectively supplement real data, especially in cases where labeled data is scarce or difficult to obtain. Through a series of augmentation experiments, we validate that our generated traffic samples can substantially improve classification performance, particularly for minority classes. By enriching the training data for underrepresented traffic categories, the model becomes more robust in accurately classifying diverse traffic patterns.

Despite the promising results, our research faces several limitations as outlined above. We aim to address these issues in our future work, such as implementing more detailed field segmentation and generation constraints for protocols, and optimizing resource consumption. We hope to conduct more in-depth investigations to explore the broader applicability of our model across various network security challenges.





\bibliographystyle{IEEEtran}
\bibliography{ref}

\end{document}